\documentclass[prd,preprint,superscriptaddress]{revtex4}

\usepackage{amsmath}
\usepackage{amssymb}
\usepackage{amsfonts}
\usepackage{mathrsfs} 
\usepackage{mathtools}

\usepackage{bm}
\usepackage{xargs}
\usepackage{longtable}
\usepackage[hidelinks]{hyperref}
\usepackage{tabularx}
\usepackage{tikz-cd} 

\newcommand*\rot{\rotatebox{-90}}

\renewcommand{\L}[1]{\ensuremath{\lambda_{#1}}}
\newcommand{\Ld}[2]{\ensuremath{\lambda_{#1}^{#2}}}
\newcommand{\Ldd}[2]{\ensuremath{\lambda_{#1}^{#2}}}
\newcommand{\LL}[2]{\ensuremath{\lambda_{\vphantom{#2}#1}{}_{\vphantom{#1}#2}}}
\newcommand{\XX}[2]{\ensuremath{\xi_{\vphantom{#2}#1}{}_{\vphantom{#1}#2}}}
\newcommand{\TT}[2]{\ensuremath{\theta_{\vphantom{#2}#1}{}_{\vphantom{#1}#2}}}
\newcommand{\LLd}[3]{\ensuremath{\lambda_{\vphantom{#2}#1}^{\vphantom{#3}}{}_{#2}^{#3}}}
\newcommand{\XXd}[3]{\ensuremath{\xi_{\vphantom{#2}#1}^{\vphantom{#3}}{}_{#2}^{#3}}}
\newcommand{\LLdd}[3]{\ensuremath{\lambda_{\vphantom{#2}#1}^{\vphantom{#3}}{}_{#2}^{#3}}}
\newcommand{\LdLd}[4]{\ensuremath{\lambda_{\vphantom{#3}#1}^{#2}{}_{\vphantom{#1}#3}^{#4}}}
\newcommand{\LdLdd}[4]{\ensuremath{\lambda_{#1}^{\vphantom{#4}#2}{}_{#3}^{\vphantom{#2}#4}}}
\newcommand{\LddLdd}[4]{\ensuremath{\lambda_{#1}^{\vphantom{#4}#2}{}_{#3}^{\vphantom{#2}#4}}}

\newcommandx{\E}[3][3=]{\ensuremath{E^{#1}{}_{#2}{}_{#3}}}
\newcommandx{\F}[4][4=]{\ensuremath{F^{#1}{}_{#2}{}^{#3}{}_{#4}}}
\newcommandx{\W}[5][5=]{\ensuremath{W^{#1}{}_{#2\,#3}{}^{#4}{}_{#5}}}
\newcommandx{\Wi}[5][5=]{\ensuremath{W^{#1}{}_{#2\,#3}{}^{#4}{}_{#5}}}
\newcommandx{\BC}[5][5=]{\ensuremath{B^{#1}{}_{#2\,#3}{}^{#4}{}_{#5}}}
\newcommandx{\Bi}[5][5=]{\ensuremath{B^{#1}{}_{#2\,#3}{}^{#4}{}_{#5}}}
\renewcommandx{\H}[4][4=]{\ensuremath{H^{#1}{}_{#2\,#3}{}_{#4}}}
\newcommandx{\Hi}[4][4=]{\ensuremath{H^{#1}{}_{#2\,#3}{}_{#4}}}

\newcommandx{\U}[3][3=]{\ensuremath{U^{#1\,#2}{}_{#3}}}
\newcommandx{\V}[3][3=]{\ensuremath{V^{#1\,#2}{}_{#3}}}
\newcommandx{\MC}[3][3=]{\ensuremath{M^{#1\,#2}{}_{#3}}}
\newcommandx{\Q}[4][4=]{\ensuremath{Q_{#1}{}^{#2\,#3}{}_{#4}}}
\newcommandx{\Qi}[4][4=]{\ensuremath{Q_{#1}{}^{#2\,#3}{}_{#4}}}
\newcommandx{\T}[3][3=]{\ensuremath{T^{#1\,#2}{}_{#3}}}

\newcommand{\I}{\mathcal{I}}

\newcommand{\Ii}[2]{\ensuremath{\mathcal{I}^{\overline{#1}}{}_{#2}}}
\newcommand{\Iii}[2]{\ensuremath{\mathcal{I}^{\overline{\overline{#1}}}{}^{#2}}}
\newcommand{\Iiii}[3]{\ensuremath{\mathcal{I}^{\overline{\overline{\overline{#1}}}}{}_{#2}{}^{#3}}}
\renewcommand{\i}[2]{\ensuremath{\mathcal{I}^{#1}{}_{\overline{#2}}}}
\newcommand{\ii}[2]{\ensuremath{\mathcal{I}_{#1}{}_{\overline{\overline{#2}}}}}
\newcommand{\iii}[3]{\ensuremath{\mathcal{I}^{#1}{}_{#2}{}_{\overline{\overline{\overline{#3}}}}}}

\renewcommand{\r}[1]{\ensuremath{\left(#1\right)}}
\newcommand{\e}[1]{\ensuremath{\left[ #1 \right]}}

\newcommand{\rp}[1]{\ensuremath{\left( #1 \right.}}
\newcommand{\pr}[1]{\ensuremath{\left. #1 \right)}}

\newcommandx{\adots}[2][1=1]{\ensuremath{\alpha_{#1}\ldots\alpha_{#2}}}
\newcommandx{\bdots}[2][1=1]{\ensuremath{\beta_{#1}\ldots\beta_{#2}}}
\newcommandx{\mdots}[2][1=1]{\ensuremath{\mu_{#1}\ldots\mu_{#2}}}
\renewcommand{\a}[1]{\ensuremath{\alpha_{#1}}}
\renewcommand{\b}[1]{\ensuremath{\beta_{#1}}}
\newcommand{\m}[1]{\ensuremath{\mu_{#1}}}
\newcommand{\s}[1]{\ensuremath{\sigma_{#1}}}
\renewcommand{\t}[1]{\ensuremath{\tau_{#1}}}
\newcommand{\Aa}{\overline{A}}
\newcommand{\Ab}{\overline{\overline{A}}}
\newcommand{\Ac}{\overline{\overline{\overline{A}}}}
\newcommand{\Ba}{\overline{B}}

\newcommandx{\p}[2][2=]{\ensuremath{\varphi^{\overline{#1}}{}_{#2}}}
\newcommandx{\pp}[2][2=]{\ensuremath{\varphi^{\overline{\overline{#1}}}{}_{#2}}}
\newcommandx{\ppp}[2][2=]{\ensuremath{\varphi^{\overline{\overline{\overline{#1}}}}{}_{#2}}}

\newcommandx{\cP}[2][2=]{\ensuremath{\Phi^{#1}{}_{#2}}}
\renewcommandx{\P}[2][2=]{\ensuremath{\varphi^{#1}{}_{#2}}}

\newcommandx{\Cd}[3][3=]{\ensuremath{C_{#1|#2}{}^{#3}}}
\newcommandx{\Cid}[3][3=]{\ensuremath{C_{#1|#2}{}^{#3}}}

\newcommand{\kd}[2]{\ensuremath{\delta^{#1}_{#2}}}
\newcommand{\gu}[1]{\ensuremath{\gamma^{#1}}}
\newcommand{\gd}[1]{\ensuremath{\gamma_{#1}}}

\newcommand{\eps}[3]{\ensuremath{\epsilon{#1}{}{#2}{}{#3}}}
\newcommand{\euuu}[3]{\ensuremath{\epsilon^{#1#2#3}}}
\newcommand{\euud}[3]{\ensuremath{\epsilon^{#1#2}{}_{#3}}}
\newcommand{\eudd}[3]{\ensuremath{\epsilon^{#1}{}_{#2#3}}}

\newcommand{\eudu}[3]{\ensuremath{\epsilon^{#1}{}_{#2}{}^{#3}}}

\newcommand{\eddd}[3]{\ensuremath{\epsilon_{#1#2#3}}}

\newcommand{\lb}[1]{\ensuremath{\lambda_{\overline{#1}}}}
\newcommand{\lbb}[1]{\ensuremath{\lambda_{\overline{\overline{#1}}}}}
\newcommand{\lbbb}[1]{\ensuremath{\lambda_{\overline{\overline{\overline{#1}}}}}}
\newcommand{\lbd}[2]{\ensuremath{\lambda_{\overline{#1}}^{#2}}}
\newcommand{\lbbd}[2]{\ensuremath{\lambda_{\overline{\overline{#1}}}^{#2}}}

\newcommand{\lbdd}[2]{\ensuremath{\lambda_{\overline{#1}}^{#2}}}
\newcommand{\lbbdd}[2]{\ensuremath{\lambda_{\overline{\overline{#1}}}^{#2}}}
\newcommand{\lbbbdd}[2]{\ensuremath{\lambda_{\overline{\overline{\overline{#1}}}}^{#2}}}

\newcommand{\lbB}[2]{\ensuremath{\lambda_{\overline{#1}\,{\overline{#2}}}}}
\newcommand{\lbBB}[2]{\ensuremath{\lambda_{\overline{#1}\,{\overline{\overline{#2}}}}}}
\newcommand{\lbBBB}[2]{\ensuremath{\lambda_{\overline{#1}\,{\overline{\overline{\overline{#2}}}}}}}
\newcommand{\lbbBB}[2]{\ensuremath{\lambda_{\overline{\overline{#1}}\,{\overline{\overline{#2}}}}}}
\newcommand{\lbbBBB}[2]{\ensuremath{\lambda_{\overline{\overline{#1}}\,{\overline{\overline{\overline{#2}}}}}}}
\newcommand{\lbbbBBB}[2]{\ensuremath{\lambda_{\overline{\overline{\overline{#1}}}\,{\overline{\overline{\overline{#2}}}}}}}

\newcommand{\xbBB}[2]{\ensuremath{\xi_{\overline{#1}\,{\overline{\overline{{#2}}}}}}}
\newcommand{\xbBBB}[2]{\ensuremath{\xi_{\overline{#1}\,{\overline{\overline{\overline{#2}}}}}}}
\newcommand{\xbbB}[2]{\ensuremath{\xi_{\overline{\overline{#1}}\,{\overline{#2}}}}}

\newcommand{\xbbBBB}[2]{\ensuremath{\xi_{\overline{\overline{#1}}\,{\overline{\overline{\overline{#2}}}}}}}
\newcommand{\xbbbB}[2]{\ensuremath{\xi_{\overline{\overline{\overline{#1}}}\,{\overline{#2}}}}}
\newcommand{\xbbbBB}[2]{\ensuremath{\xi_{\overline{\overline{\overline{#1}}}\,{\overline{\overline{#2}}}}}}

\newcommand{\tbB}[2]{\ensuremath{\theta_{\overline{#1}\,{\overline{#2}}}}}
\newcommand{\tbBB}[2]{\ensuremath{\theta_{\overline{#1}\,{\overline{\overline{#2}}}}}}
\newcommand{\tbBBB}[2]{\ensuremath{\theta_{\overline{#1}\,{\overline{\overline{\overline{#2}}}}}}}
\newcommand{\tbbBB}[2]{\ensuremath{\theta_{\overline{\overline{#1}}\,{\overline{\overline{#2}}}}}}
\newcommand{\tbbBBB}[2]{\ensuremath{\theta_{\overline{\overline{#1}}\,{\overline{\overline{\overline{#2}}}}}}}
\newcommand{\tbbbBBB}[2]{\ensuremath{\theta_{\overline{\overline{\overline{#1}}}\,{\overline{\overline{\overline{#2}}}}}}}

\newcommand{\lbBd}[3]{\ensuremath{\lambda_{\overline{#1}\,{\overline{#2}}}^{\hphantom{\;#1}#3}}}
\newcommand{\lbBBd}[3]{\ensuremath{\lambda_{\overline{#1}\,{\overline{\overline{#2}}}}^{\hphantom{\;#1}#3}}}
\newcommand{\lbBBBd}[3]{\ensuremath{\lambda_{\overline{#1}\,{\overline{\overline{\overline{#2}}}}}^{\hphantom{\;#1}#3}}}
\newcommand{\lbbBd}[3]{\ensuremath{\lambda_{\overline{\overline{#1}}\,{\overline{#2}}}^{\hphantom{\;#1}#3}}}
\newcommand{\lbbBBd}[3]{\ensuremath{\lambda_{\overline{\overline{#1}}\,{\overline{\overline{#2}}}}^{\hphantom{\;#1}#3}}}
\newcommand{\lbbBBBd}[3]{\ensuremath{\lambda_{\overline{\overline{#1}}\,{\overline{\overline{\overline{#2}}}}}^{\hphantom{\;#1}#3}}}
\newcommand{\lbbbBd}[3]{\ensuremath{\lambda_{\overline{\overline{\overline{#1}}}\,{\overline{#2}}}^{\hphantom{\;#1}#3}}}
\newcommand{\lbbbBBd}[3]{\ensuremath{\lambda_{\overline{\overline{\overline{#1}}}\,{\overline{\overline{#2}}}}^{\hphantom{\;#1}#3}}}
\newcommand{\lbbbBBBd}[3]{\ensuremath{\lambda_{\overline{\overline{\overline{#1}}}\,{\overline{\overline{\overline{#2}}}}}^{\hphantom{\;#1}#3}}}

\newcommand{\xbBBd}[3]{\ensuremath{\xi_{\overline{#1}\,{\overline{\overline{#2}}}}^{\hphantom{\;#1}#3}}}
\newcommand{\xbBBBd}[3]{\ensuremath{\xi_{\overline{#1}\,{\overline{\overline{\overline{#2}}}}}^{\hphantom{\;#1}#3}}}
\newcommand{\xbbBd}[3]{\ensuremath{\xi_{\overline{\overline{#1}}\,{\overline{#2}}}^{\hphantom{\;#1}#3}}}

\newcommand{\xbbBBBd}[3]{\ensuremath{\xi_{\overline{\overline{#1}}\,{\overline{\overline{\overline{#2}}}}}^{\hphantom{\;#1}#3}}}
\newcommand{\xbbbBd}[3]{\ensuremath{\xi_{\overline{\overline{\overline{#1}}}\,{\overline{#2}}}^{\hphantom{\;#1}#3}}}
\newcommand{\xbbbBBd}[3]{\ensuremath{\xi_{\overline{\overline{\overline{#1}}}\,{\overline{\overline{#2}}}}^{\hphantom{\;#1}#3}}}

\newcommand{\lbBdd}[3]{\ensuremath{\lambda_{\overline{#1}\,{\overline{#2}}}^{\hphantom{\,#1}#3}}}
\newcommand{\lbBBdd}[3]{\ensuremath{\lambda_{\overline{#1}\,{\overline{\overline{#2}}}}^{\hphantom{\,#1}#3}}}
\newcommand{\lbBBBdd}[3]{\ensuremath{\lambda_{\overline{#1}\,{\overline{\overline{\overline{#2}}}}}^{\hphantom{\,#1}#3}}}
\newcommand{\lbbBdd}[3]{\ensuremath{\lambda_{\overline{\overline{#1}}\,{\overline{#2}}}^{\hphantom{\,#1}#3}}}
\newcommand{\lbbBBdd}[3]{\ensuremath{\lambda_{\overline{\overline{#1}}\,{\overline{\overline{#2}}}}^{\hphantom{\,#1}#3}}}
\newcommand{\lbbBBBdd}[3]{\ensuremath{\lambda_{\overline{\overline{#1}}\,{\overline{\overline{\overline{#2}}}}}^{\hphantom{\,#1}#3}}}
\newcommand{\lbbbBdd}[3]{\ensuremath{\lambda_{\overline{\overline{\overline{#1}}}\,{\overline{#2}}}^{\hphantom{\,#1}#3}}}
\newcommand{\lbbbBBdd}[3]{\ensuremath{\lambda_{\overline{\overline{\overline{#1}}}\,{\overline{\overline{#2}}}}^{\hphantom{\,#1}#3}}}
\newcommand{\lbbbBBBdd}[3]{\ensuremath{\lambda_{\overline{\overline{\overline{#1}}}\,{\overline{\overline{\overline{#2}}}}}^{\hphantom{\,#1}#3}}}

\newcommand{\lbdBd}[4]{\ensuremath{\lambda_{\overline{#1}\,\overline{#3}}^{\,#2\;#4}}}
\newcommand{\lbdBBd}[4]{\ensuremath{\lambda_{\overline{#1}\,\overline{\overline{#3}}}^{\,#2\;#4}}}
\newcommand{\lbdBBBd}[4]{\ensuremath{\lambda_{\overline{#1}\,\overline{\overline{\overline{#3}}}}^{\,#2\;#4}}}
\newcommand{\lbbdBBd}[4]{\ensuremath{\lambda_{\overline{\overline{#1}}\,\overline{\overline{#3}}}^{\,#2\;#4}}}
\newcommand{\lbbdBBBd}[4]{\ensuremath{\lambda_{\overline{\overline{#1}}\,\overline{\overline{\overline{#3}}}}^{\,#2\;#4}}}
\newcommand{\lbbbdBBBd}[4]{\ensuremath{\lambda_{\overline{\overline{\overline{#1}}}\,\overline{\overline{\overline{#3}}}}^{\,#2\;#4}}}

\newcommand{\AxbBB}[2]{\ensuremath{\r{}}}

\newcommand{\AxbbB}[2]{\ensuremath{\r{}}}

\newcommand{\AlbBBBd}[3]{\ensuremath{\r{}}}

\newcommand{\AlbbBBBd}[3]{\ensuremath{\r{}}}
\newcommand{\AlbbbBd}[3]{\ensuremath{\r{}}}
\newcommand{\AlbbbBBd}[3]{\ensuremath{\r{}}}

\newcommand{\AxbBBd}[3]{\ensuremath{\r{}}}

\newcommand{\AxbbBd}[3]{\ensuremath{\r{}}}

\newcommand{\ExbBB}[2]{\ensuremath{\e{}}}

\newcommand{\ExbbB}[2]{\ensuremath{\e{}}}

\newcommand{\ElbBBBd}[3]{\ensuremath{\e{}}}

\newcommand{\ElbbBBBd}[3]{\ensuremath{\e{}}}
\newcommand{\ElbbbBd}[3]{\ensuremath{\e{}}}
\newcommand{\ElbbbBBd}[3]{\ensuremath{\e{}}}

\newcommand{\ExbBBd}[3]{\ensuremath{\e{}}}

\newcommand{\ExbbBd}[3]{\ensuremath{\e{}}}

\renewcommand{\O}[1]{\ensuremath{\mathcal{O}(#1)}}

\newcommand{\dofa}[2][]{
	\def\tmp{#1}\ifx\tmp\empty 
		\overline{\varphi}{}^{#2}
	\else 
		\overline{\varphi}{}^{#2}{}_{,#1}
	\fi
}
\newcommand{\dofb}[2][]{
	\def\tmp{#1}\ifx\tmp\empty 
		\overline{\overline{\varphi}}{}_{#2}
	\else 
		\overline{\overline{\varphi}}{}_{#2}{}_{,#1}
	\fi
}
\newcommand{\dofc}[3][]{
	\def\tmp{#1}\ifx\tmp\empty 
		\overline{\overline{\overline{\varphi}}}{}^{#2}{}_{#3}
	\else 
		\overline{\overline{\overline{\varphi}}}{}^{#2}{}_{#3}{}_{,#1}
	\fi
}

\newcommand{\vela}[2][]{
	\def\tmp{#1}\ifx\tmp\empty 
		\overline{k}{}^{#2}
	\else 
		\overline{k}{}^{#2}{}_{,#1}
	\fi
}
\newcommand{\velb}[2][]{
	\def\tmp{#1}\ifx\tmp\empty 
		\overline{\overline{k}}{}_{#2}
	\else 
		\overline{\overline{k}}{}_{#2}{}_{,#1}
	\fi
}
\newcommand{\velc}[3][]{
	\def\tmp{#1}\ifx\tmp\empty 
		\overline{\overline{\overline{k}}}{}^{#2}{}_{#3}
	\else 
		\overline{\overline{\overline{k}}}{}^{#2}{}_{#3}{}_{,#1}
	\fi
}

\begin{document}
\title{Gravitational closure of weakly birefringent electrodynamics}
\author{Jonas Schneider}
\author{Frederic P. Schuller\footnote{Corresponding author. Email: fps@aei.mpg.de}}
\affiliation{Department of Physics, Friedrich-Alexander University Erlangen-Nuremberg, Staudtstr.~7, 91058 Erlangen, Germany}
\author{Nadine Stritzelberger}
\affiliation{Newnham College, University of Cambridge, Sidgwick Av., CB3 9DF Cambridge, United Kingdom}
\author{Florian Wolz}
\affiliation{Department of Physics, Friedrich-Alexander University Erlangen-Nuremberg, Staudtstr.~7, 91058 Erlangen, Germany}

\begin{abstract}
We derive the gravitational dynamics 
of the tensorial geometry which underlies the most general linear theory of electrodynamics  that features  weak birefringence in vacuo. This derivation is performed by way of gravitational closure, which is a mechanism that employs the causal structure of any canonically quantizable matter dynamics on some tensorial spacetime geometry in order to derive canonical dynamics for the latter. The resulting eleven-parameter family of weak gravitational field equations allows to predict where vacuum birefringence will occur, if there is any.
\end{abstract}

\maketitle

\tableofcontents

\newpage
\section{Introduction}
The motion of matter fields crucially depends on the coefficient functions in their equations of motion. In order to close these equations, one thus needs to provide dynamics for their coefficient functions as well.
The simplest example is a Klein-Gordon equation
\begin{equation}\label{KGeqn}
g^{ab}\partial_a\partial_b\phi - \frac{1}{2}g^{mn} g^{as}(\partial_m g_{sn} + \partial_n g_{ms} - \partial_s g_{mn}) \partial_a \phi - m^2 \phi = 0
\end{equation}
for a scalar matter field $\phi$, where the coefficient functions are provided by the components of a Lorentzian metric $g$. It is usually posited that for physical reasons, one must choose the Einstein-Hilbert dynamics 
\begin{equation}\label{luckybastard}
R_{ab}[g] - \frac{1}{2} (R[g]-2\Lambda) g_{ab}  = 8\pi G \, T_{ab}[\phi;g)
\end{equation}
in order to close these matter equations, or any other in the standard model of particle physics, with the constants $\Lambda$ and $G$ being determined by experiment. 

There is a way to obtain the equations (\ref{luckybastard}) in a constructive way directly from the matter equations (\ref{KGeqn}), under the combined assumptions of the classic paper \cite{HKT} and the more recent work \cite{RRS}. The underlying mechanism is quite simple. Requiring the dynamics for the coefficients to enjoy a canonical evolution that starts and ends on the same initial value hypersurfaces as the given matter dynamics, the admissible spectrum of diffeomorphism-invariant coefficient dynamics is severely constrained. Indeed, one finds that the dynamics for the coefficients must arise as the solution to an immutable set of {\it gravitational closure equations} \cite{WittePhD}, which convert input from the given matter dynamics into output that determines the required action for the coefficients. In physics parlance, this mechanism amounts to a  constructive derivation of gravitational dynamics from specified matter dynamics.    

The very same gravitational closure mechanism is applicable to any canonically quantizable matter action on any tensorial background \cite{DSSW}. This article is now concerned with the first explicit execution of this mechanism for one specific matter theory beyond the standard model, namely the birefringent electrodynamics 
\begin{equation}\label{GLED}
S_\textrm{\tiny matter}[A;G) = -\frac{1}{8} \int_M d^4x \,\omega_G(x) \,G^{abcd}(x) F_{ab}(x) F_{cd}(x)\,,
\end{equation}
which has been extensively studied as a classical \cite{HehlBook,Rubilar,SWW} and quantum matter theory \cite{SR,Pfeifer} in the context of standard model extensions \cite{Kostelecky,theentireposse} and optics \cite{Perlick} on a four-dimensional smooth manifold $M$. The employed geometric structure, and thus the coefficient functions of the associated matter field equations for the covector field $A$, is given by a fourth-rank tensor field $G^{abcd}$ 
and the induced weight-one scalar density $\omega_G:=24(\epsilon_{abcd} G^{abcd})^{-1}$.

Gravitational closure equations for these birefringent electrodynamics, whose solution provides gravitational dynamics for the tensor field $G$, have been set up in \cite{DSSW}, but are prohibitively difficult to solve exactly. But for many practical purposes, such exact gravitational field equations are not needed in the first place. For even if they are available, like the Einstein equations for a Lorentzian metric, they are then solved either by imposing symmetry assumptions or by perturbative techniques. While the imposition of symmetry conditions at the level of the closure equations comes with caveats concerning symmetric criticality \cite{Palais,moderncriticality}, a perturbation ansatz for the background geometry can be made, without difference, at any stage of the closure procedure: one may follow any path in the diagram
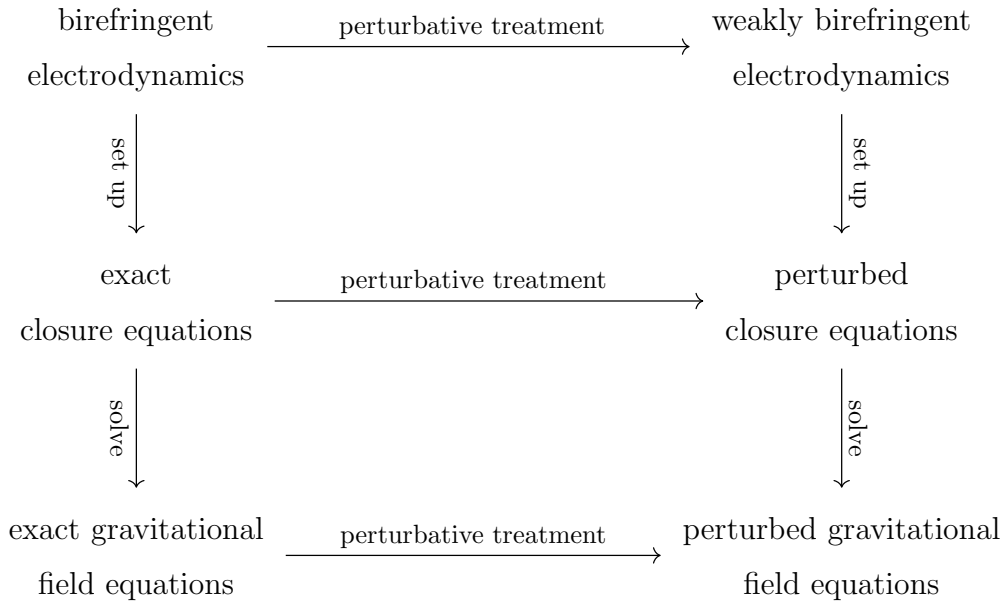
\[\begin{tikzcd}[row sep=1.6cm, column sep=5cm]
\begin{array}{c}
\textrm{birefringent}\\ 
\textrm{electrodynamics}
\end{array} 
\arrow{r}{\text{\footnotesize perturbative treatment \,}} 
\arrow[swap]{d}{\rot{\text{\footnotesize set up}}} 
& 
\begin{array}{c}
\textrm{weakly birefringent}\\
\textrm{electrodynamics}
\end{array} 
\arrow{d}{\rot{\text{\footnotesize set up}}} \\
\begin{array}{c}
\textrm{exact}\\ 
\textrm{closure equations}
\end{array} 
\arrow{r}{\text{\footnotesize perturbative treatment \,\,\, }} 
\arrow[swap]{d}{\rot{\text{\footnotesize solve}}} 
& 
\begin{array}{c}
\textrm{perturbed}\\ 
\textrm{closure equations}
\end{array} 
\arrow{d}{\rot{\text{\footnotesize solve}}} \\
\begin{array}{c}
\textrm{exact gravitational}\\ 
\textrm{field equations}
\end{array} 
\arrow{r}{\text{\footnotesize perturbative treatment}} 
& 
\begin{array}{c}
\textrm{perturbed gravitational}\\
\textrm{field equations}
\end{array}
\end{tikzcd}
\]

In this article, we employ a set-up for the closure equations of birefringent electrodynamics that is tailored to their perturbative solution. In particular, we employ a practical parametrization of the canonical configuration degrees of freedom for the fourth-rank geometry $G$ that describe small perturbations around a flat background and calculate the input coefficients, which are needed to set up the closure equations, directly from weakly birefringent electrodynamics in section \ref{sec_closureeqn}. We solve the perturbed  closure equations to obtain the gravitational Lagrangian for $G$ to quadratic order in section \ref{sec_perteval}, display the linearized gravitational field equations in a convenient gauge in section \ref{sec:gaugedeom}, and conclude in section \ref{sec_conclusions}.

\newpage
\section{Closure equations for birefringent electrodynamics}\label{sec_closureeqn}
We set up the gravitational closure equations for weakly birefringent electrodynamics in this section. In order to keep this article technically self-contained, the first two subsections review how the canonical geometry underlying birefringent electrodynamics is obtained. Starting with the third subsection, we then turn to new results. In particular, we employ a parametrization of the canonical geometry that is is tailored to the needs of the perturbative treatment undertaken in the remainder of this work. The central result of this section are the input coefficients that define the gravitational closure equations for weakly birefringent electrodynamics, which we  will then solve in the next section.

\subsection{Birefringent electrodynamics and the underlying spacetime $(M,G,P_G)$}
The physical starting point, and single input from which all results obtained in this article follow, is the electrodynamic theory  for an abelian gauge potential $A$ given by the action (\ref{GLED}).
This matter theory is the most general theory of electrodynamics with a tensorial background geometry that still features a classical superposition principle \cite{SWW}. Vacuum birefringence is not excluded a priori, in contrast to the limit case of Maxwell electrodynamics, to which the equations of motion associated with the above action reduce if the background geometry is induced from a metric $g$ by virtue of  $G_g^{abcd} := g^{ac}g^{bd} - g^{ad}g^{bc} - \sqrt{-\det g^{\cdot\cdot}} \epsilon^{abcd}$. Considering only the components of $G$ that contribute to the above action, we may assume without loss of generality that the geometric tensor features the algebraic symmetries
$$
G^{abcd} = G^{cdab} \quad \textrm{ and } \quad G^{abcd} = - G^{bacd}\,.
$$

The principal tensor $P_G$ of the equations of motion associated with the above action is  found \cite{Rubilar} to be given by
\begin{equation}\label{Ppoly}
P_G^{abcd} = -\frac{1}{24} \omega_G^2 \epsilon_{mnpq} \epsilon_{rstu} G^{mnr(a} G^{b|sp|c} G^{d)qtu}
\end{equation}
in terms of the spacetime geometry $G$ and the unique totally antisymmetric tensor density of weight $-1$ that is normalized such that $\epsilon_{0123}=1$.  Since the dynamics (\ref{GLED}) are canonically quantizable \cite{SR,Pfeifer}, the triple $(M,G,P_G)$ contains the complete kinematical theory \cite{RRS} required to physically interpret the geometry $G$ and to set up the gravitational closure equations, whose  solution then provides the dynamics of the geometry according to \cite{DSSW}.  

\subsection{Induced geometry and canonical geometry}
\label{sec_spatial_tensor_fields}
Employing the projections from section II.B of \cite{DSSW}, which are induced by a family of embedding maps  $X_t: \Sigma \to M$ that foliate the manifold $(M,G,P_G)$ into initial data surfaces $X_t(\Sigma)$ for the matter dynamics (\ref{GLED}), one obtains an induced geometry on the three-manifold $\Sigma$ given by the one-parameter family of tensor fields 
\begin{eqnarray*}
\mathbf{g}^{\beta\delta}[X](t,y) &:=& - G\left(\epsilon^0_t(y) , \epsilon_t^{\beta}(y), \epsilon^0_t(y) , \epsilon_t^{\delta}(y) \right)\,, \\
\mathbf{g}_{\alpha\beta}[X](t,y) &:=& \frac{1}{4} \frac{\epsilon_{\alpha\mu\nu}}{\sqrt{\text{det}\,\mathbf{g}^{\cdot\cdot}}}\frac{\epsilon_{\beta\rho\sigma}}{\sqrt{\text{det}\,\mathbf{g}^{\cdot\cdot}}} \,  G \left(\epsilon_t^{\mu}(y), \epsilon_t^{\nu}(y), \epsilon_t^{\rho}(y), \epsilon_t^{\sigma}(y) \right) \\
\mathbf{g}^{\alpha}{}_{\beta}[X](t,y) &:=& \frac{1}{2} \frac{\epsilon_{\beta\gamma\delta}}{\sqrt{\text{det}\,\mathbf{g}^{\cdot\cdot}}} \,   G\left( \epsilon^0_t(y), \epsilon_t^{\alpha}(y), \epsilon_t^{\gamma}(y), \epsilon_t^{\delta}(y)\right) - \delta^{\alpha}_{\beta}\,,
\end{eqnarray*}
which present the so-called induced geometry in the terminology of \cite{DSSW}.
Since the principal tensor $P_G$ is determined entirely in terms of the spacetime geometry, it is likewise projected to several one-parameter families of tensor fields on $\Sigma$, which can  then be expressed solely in terms of the induced geometry $\mathbf{g}^\mathscr{A}$. Out of these, only the components  
\begin{eqnarray*}
\mathbf{p}^{\alpha\beta}[X](y) &=& \frac{1}{6}( \mathbf{g}^{\alpha\gamma} \mathbf{g}^{\beta\delta} \mathbf{g}_{\gamma\delta} - \mathbf{g}^{\alpha\beta}\mathbf{g}^{\gamma\delta}\mathbf{g}_{\gamma\delta} - 2 \, \mathbf{g}^{\alpha\beta} \mathbf{g}^{\gamma}{}_{\delta} \mathbf{g}^{\delta}{}_{\gamma} + 3 \, \mathbf{g}^{\gamma\delta} \mathbf{g}^{\alpha}{}_{\gamma} \mathbf{g}^{\beta}{}_{\delta}   ) \,
\end{eqnarray*}
appear in the gravitational closure equations and are thus the only ones we need to consider. 
By construction, the induced geometry satisfies the overall non-linear  frame conditions
$$\mathbf{g}\,{}^{\mu [\alpha} \mathbf{g}\,{}^{\beta]}{}_{\mu} = 0 \qquad \text{and} \qquad \mathbf{g}\,{}^{\alpha}{}_{\alpha} = 0$$
and the additional linear symmetry conditions
$$\mathbf{g}\,{}^{[\alpha\beta]} = 0\, \qquad \text{ and} \qquad \mathbf{g}\,{}_{[\alpha\beta]} = 0\,,$$
as described in section III.B of \cite{DSSW}.

Now we change perspective from the induced geometry (where the spacetime geometry $G$ is considered as primary and the induced geometry as secondary) to the canonical geometry (where this hierarchy is inverted). This is done by installing 
three tensor fields 
$$g{}^{\alpha\beta}\,, \qquad g{}_{\,\alpha\beta}\,, \qquad g\,{}^\alpha{}_\beta$$
on $\Sigma$ that mimic the three families of induced tensor fields on $\Sigma$,
but are no longer considered to arise as projections of the spacetime tensor field $G$. As explained in section III.B of \cite{DSSW}, the frame and symmetry properties of the induced geometry must thus be restored explicitly by demanding that 
$$ 
g{}^{\mu [\alpha} g{}^{\beta]}{}_{\mu} = 0\,, \qquad  g{}^{\alpha}{}_{\alpha} = 0 \qquad \text{and} \qquad  g{}^{[\alpha\beta]} = 0\,, \qquad g{}_{[\alpha\beta]} = 0\,.
$$
Taking into account the symmetry constraints, this amounts to 17 independent initial data surface field components, which together with the four frame conditions employed in the projection, restore the total of 21 independent components  featured by the spacetime tensor $G$ at each point in spacetime. 

\subsection{Parametrization and configuration variables}
\label{sec_perturbative_parametrization}
The frame conditions for the screen manifold fields $g^{\alpha\beta}$, $g_{\alpha\beta}$, $g^\alpha{}_\beta$ derived in the previous section include a set of non-linear equations.
Thus we require the parametrization technology developed in section III.C of \cite{DSSW} and need to choose suitable parametrization maps $\widehat g^\mathscr{A}$ that generate the canoncial geometry $g^\mathscr{A}:=(g^{\alpha\beta}, g_{\alpha\beta}, g^{\alpha}{}_\beta)$ in terms of 17 unconstrained configuration variables $\varphi^1$, $\dots$, $\varphi^{17}$ such that all frame and symmetry conditions are met. 

Since this article is concerned with geometries $G^{abcd}$ that present perturbations around Minkowski spacetime $\eta$, we  choose a parametrization directly in terms of the perturbative degrees of freedom in the presence of a flat Euclidean background metric $\gamma^{\cdot\cdot}$ on $\Sigma$, namely
\begin{eqnarray*}
\widehat g^{\alpha\beta}(\varphi) &:=&  \gamma^{\alpha\beta} + \I^{\alpha\beta}{}_{\Aa} \varphi^{\Aa} \,,\\
\widehat g_{\alpha\beta}(\varphi)  &:=&  \gamma_{\alpha\beta} + \I_{\alpha\beta}{}_{\Ab} \varphi^{\Ab} \,,\\
\widehat g^{\alpha}{}_{\beta}(\varphi) &:=&  \I^{\alpha}{}_{\beta}{}_{\Ac} \varphi^{\Ac} + f^\alpha{}_\beta(\varphi)\,
\end{eqnarray*}
where the endomorphism $f^\alpha{}_\beta$ appearing in the last definition is required to be $\gamma$-antisymmetric and tracefree \cite{Reiss}, and the three intertwiners are defined as
$$
\I^{\alpha\beta}{}_{\Aa} = \tfrac{1}{\sqrt{2}} \begin{psmallmatrix} \sqrt{2} & 0 & 0 & 0 & 0 & 0 \\
0 & 1 & 0 & 0 & 0 & 0 \\
0 & 0 & 1 & 0 & 0 & 0 \\
0 & 1 & 0 & 0 & 0 & 0 \\
0 & 0 & 0 & \sqrt{2} & 0 & 0 \\
0 & 0 & 0 & 0 & 1 & 0 \\
0 & 0 & 1 & 0 & 0 & 0 \\
0 & 0 & 0 & 0 & 1 & 0 \\
0 & 0 & 0 & 0 & 0 & \sqrt{2} \end{psmallmatrix}^{\!\!\!\alpha\beta}_{\,\,\,\Aa} 
 ,\,\,
\I_{\alpha\beta}{}_{\Ab} = \tfrac{1}{\sqrt{2}} \begin{psmallmatrix} \sqrt{2} & 0 & 0 & 0 & 0 & 0 \\
0 & 1 & 0 & 0 & 0 & 0 \\
0 & 0 & 1 & 0 & 0 & 0 \\
0 & 1 & 0 & 0 & 0 & 0 \\
0 & 0 & 0 & \sqrt{2} & 0 & 0 \\
0 & 0 & 0 & 0 & 1 & 0 \\
0 & 0 & 1 & 0 & 0 & 0 \\
0 & 0 & 0 & 0 & 1 & 0 \\
0 & 0 & 0 & 0 & 0 & \sqrt{2} \end{psmallmatrix}^{\!\!\!\alpha\beta}_{\,\,\,\Ab} 
,\,\,
\I^{\alpha}{}_{\beta}{}_{\Ac} = \tfrac{1}{\sqrt{2}} \begin{psmallmatrix} 1 & \frac{1}{\sqrt{3}} & 0 & 0 & 0 \\
0 & 0 & 1 & 0 & 0  \\
0 & 0 & 0 & 0 & 1  \\
0 & 0 & 1 & 0 & 0  \\
-1 & \frac{1}{\sqrt{3}} & 0 & 0 & 0  \\
0 & 0 & 0 & 1 & 0  \\
0 & 0 & 0 & 0 & 1  \\
0 & 0 & 0 & 1 & 0  \\
0 & \frac{-2}{\sqrt{3}} & 0 & 0 & 0  \end{psmallmatrix}^{\!\!\!\vphantom{I}^{\alpha}{}_{\beta}}_{\,\,\,\Ac} 
$$
with column labels $\alpha\beta$ running in the order $11,12,13,21,22,23,31,32,33$ while the rows are labeled by $\Aa=1,\dots,6$, $\Ab=7,\dots,12$ and $\Ac=13,\dots,17$, respectively. Alternatively, it is often formally convenient to employ one generic, unbarred index $A$ running over the entire range $1, \dots, 17$ and to then treat $\I^{\alpha\beta}{}_A$ as zero unless $A$ is in the range of $\Aa$, $\I_{\alpha\beta A}$ as zero unless $A$ is in the range of $\Ab$, and analogously for the third intertwiner. With this convention, the first parametrization map above, for instance, becomes $\widehat g^{\alpha\beta}(\varphi):=\gamma^{\alpha\beta}+\I^{\alpha\beta}{}_A \varphi^A$.

The said requirement on $f^\alpha{}_\beta$, which is geared at making perturbation theory on the $\gamma$ background geometry on the screen manifold as easy as possible, already uniquely fixes this endomorphism to
$$f^\alpha{}_\beta(\varphi) = \sum_{n=1}^{\infty} (-1)^{n-1}\frac{1}{2^n} \underbrace{\Big\{ \overline{\varphi}, \Big\{ \overline{\varphi}, \Big\{ \dots , \Big\{}_{n-1 \text{ curly brackets}} \overline{\varphi} \, , \Big[\, \overline{\varphi} , \overline{\overline{\overline{\varphi}}} \,\Big] \Big\} \dots \Big\} \Big\}\Big\}\,, $$
where in the commutator $[\cdot,\cdot]$ and anticommutator $\{\cdot,\cdot\}$ brackets, we employ the endomorphisms with components $\overline{\varphi}^\alpha{}_\beta := \gamma_{\beta\sigma} \I^{\alpha\sigma}{}_M \varphi^M$ and $\overline{\overline{\overline{\varphi}}}{}^\alpha{}_\beta :=  \I^\alpha{}_{\beta}{}_M \varphi^M$. 

By virtue of the thus chosen parametrization, the quadratic condition that the field $g^{\alpha}{}_{\beta}$ be $g^{\alpha\beta}$-symmetric translates into the requirement that the intertwiner $\I^{\alpha}{}_{\beta}{}_M$ be $\gamma$-symmetric and the condition that $g^{\alpha}{}_{\beta}$ be tracefree translates into the requirement that $\I^{\alpha}{}_{\beta}{}_M$ be tracefree, both of which requirements are implemented by construction of the third intertwiner above. The intertwiners $\I^{\alpha\beta}{}_M$ and $\I_{\alpha\beta}{}_M$ are symmetric due to the symmetry of the fields $g^{\alpha\beta}$ and $g_{\alpha\beta}$.
The projectors $S^{\mathscr{A}}{}_{\mathscr{B}}$,  according to which the intertwiners are constructed following the general method developed in \cite{Reiss}, are the symmetrizers
$$
S^{\alpha_1\alpha_2}{}_{\beta_1\beta_2} = \delta^{(\alpha_1}_{\beta_1}\delta^{\alpha_2)}_{\beta_2} \quad , \quad 
S_{\alpha_1\alpha_2}{}^{\beta_1\beta_2} = \delta_{(\alpha_1}^{\beta_1}\delta_{\alpha_2)}^{\beta_2} \quad \text{and} \quad
S^{\alpha_1}{}_{\alpha_2\,\beta_1}{}^{\beta_2} = \frac{1}{2}(\delta^{\alpha_1}_{\beta_1}\delta^{\beta_2}_{\alpha_2} + \gamma^{\alpha_1\beta_2}\gamma_{\alpha_2\beta_1}) - \frac{1}{3}\delta^{\alpha_1}_{\alpha_2}\delta^{\beta_2}_{\beta_1} \,.
$$
These symmetrizers implement a symmetrization to rank two tensors and the third symmetrizer moreover removes the trace. The intertwiners and their inverses satisfy the completeness relations 
$$
\I^{\mathscr{A}}{}_A \I^{A}{}_{\mathscr{B}} = S^{\mathscr{A}}{}_{\mathscr{B}} \qquad \text{and} \qquad
\I^{\mathscr{A}}{}_A \I^{B}{}_{\mathscr{A}} = \delta^B_A \,.
$$
The exact intertwiners that are defined in section III.C of \cite{DSSW}, and which are required from the next section onwards, can be expressed in terms of the three constant intertwiners introduced above. Indeed, one obtains 
$$
 \frac{\partial \widehat{\varphi}^{\overline{A}}}{\partial g^{\mathscr{A}}} \,=\,  \mathcal{I}{}^{\overline{A}}{}_{\mathscr{A}}\,, 
	\qquad
 \frac{\partial \widehat{\varphi}^{\overline{\overline{A}}}}{\partial g^{\mathscr{A}}} \,=\,  \mathcal{I}^{\overline{\overline{A}}}{}_{\mathscr{A}} \,,
	 \qquad
	\frac{\partial\widehat \varphi^{\overline{\overline{\overline{A}}}}}{\partial g^\mathscr{A}}(\varphi) = \mathcal{I}{}^{\overline{\overline{\overline{A}}}}{}_\mathscr{A}+\mathcal{O}(\varphi)\,,
$$
where only the third intertwiner is seen to differ from the constant one. These higher order modifications are needed in the next section when we calculate the input coefficients that enter the gravitational closure equations. 
 
Note that one can also find a parametrization of the tensor fields without making use of any additional structure like the flat background metric $\gamma$, as was presented in section V.C of \cite{DSSW}. One would use the parametrization given there, or any other in the same spirit,  
 when solving the gravitational closure equations exactly, rather than perturbatively.
 
\subsection{Input coefficients for the gravitational closure equations}
All preparations have now been made in order to calculate the complete information about the initially specified matter dynamics that trickles down to the gravitational closure equations for their geometric background: the input coefficients of section IV.C of \cite{DSSW}.

We only present the input coefficients up to the order that will be required by the perturbative evaluation of the closure equations in the next section. Tedious, but straightforward calculation yields
\begin{align*}
M^{\Aa \gamma} &= 2 \, \epsilon^{\mu \gamma \alpha} \, \I^{\Aa}{}_{\alpha\beta} \, \I^{\beta}{}_{\mu \, M} \, \varphi^M \\
&+ \epsilon^{\mu \gamma \alpha} \, \I^{\Aa}{}_{\alpha\beta} \, \left( \gamma_{\sigma\mu} \, \I^{\beta\tau}{}_{M} \, \I^{\sigma}{}_{\tau \, N}
- \gamma_{\sigma\mu} \, \I^{\sigma\tau}{}_{M} \, \I^{\beta}{}_{\tau \, N}
+ \gamma_{\sigma\tau} \, \I^{\sigma\tau}{}_{M} \, \I^{\beta}{}_{\mu \, N} \right) \varphi^M \varphi^N 
+ \mathcal{O}(\varphi^3) \,,\\ 
M^{\Ab \, \gamma} &= 2 \, \epsilon_{\mu\nu\alpha} \gamma^{\nu\gamma} \, \I^{\Ab \, \alpha\beta} \, \I^{\mu}{}_{\beta \, M} \, \varphi^M \\
&+ \epsilon_{\mu\nu\alpha} \, \I^{\Ab \, \alpha\beta} \, \Big( - \I^{\mu\gamma}{}_{M} \, \I^{\nu}{}_{\beta \, N}
+ \gamma^{\mu\sigma} \gamma^{\gamma\tau} \, \I_{\sigma\tau \, M} \, \I^{\nu}{}_{\beta \, N}
- \gamma^{\mu\gamma}\gamma^{\sigma\tau} \, \I_{\sigma\tau \, M} \, \I^{\nu}{}_{\beta\, N}\\
&  - \gamma^{\mu\gamma}\gamma_{\tau\beta}\, \I^{\nu\sigma}{}_{M} \, \I^{\tau}{}_{\sigma \, N}
+ \gamma^{\mu\gamma} \gamma_{\tau\beta} \, \I^{\sigma\tau}{}_{N} \, \I^{\nu}{}_{\sigma \, N} \Big) \varphi^M \varphi^N  
+ \mathcal{O}(\varphi^3) \,,\\
M^{\Ac \, \gamma} &= - \epsilon^{\mu\gamma\alpha} \I^{\Ac}{}_{\alpha}{}^{\beta} \, \I_{\mu\beta \, M} \varphi^M
- \epsilon_{\mu\nu\beta} \gamma^{\nu\gamma} \I^{\Ac}{}_{\alpha}{}^{\beta} \, \I^{\alpha\mu}{}_{M} \, \varphi^M
+ \mathcal{O}(\varphi^2)\,
\end{align*}
for the input coefficients $M^{A\gamma}$ and 
\begin{align*}
p^{\alpha\beta} &= \frac{1}{6} \big[ -2 \gamma^{\alpha\beta} - \I^{\alpha\beta}{}_M \varphi^M - \gamma^{\alpha\beta} \gamma_{\gamma\delta} \I^{\gamma\delta}{}_M \varphi^M + \gamma^{\alpha\gamma} \gamma^{\beta\delta} \I_{\gamma\delta}{}_M \varphi^M - \gamma^{\alpha\beta} \gamma^{\gamma\delta} \I_{\gamma\delta}{}_M \varphi^M \\
&+ \gamma_{\gamma\delta} \I^{\alpha\gamma}{}_M \I^{\beta\delta}{}_N \varphi^M \varphi^N
- \gamma_{\gamma\delta} \I^{\alpha\beta}{}_M \I^{\gamma\delta}{}_N \varphi^M \varphi^N
+ \gamma^{\beta\delta} \I^{\alpha\gamma}{}_M \I_{\gamma\delta}{}_N \varphi^M \varphi^N
+ \gamma^{\alpha\gamma} \I^{\beta\delta}{}_M \I_{\gamma\delta}{}_N \varphi^M \varphi^N \\
&- \gamma^{\gamma\delta} \I^{\alpha\beta}{}_M \I_{\gamma\delta}{}_N \varphi^M \varphi^N
- \gamma^{\alpha\beta} \I^{\gamma\delta}{}_M \I_{\gamma\delta}{}_N \varphi^M \varphi^N
- 2 \gamma^{\alpha\beta} \I^{\gamma}{}_{\delta}{}_M \I^{\delta}{}_{\gamma}{}_N \varphi^M \varphi^N
+ 3 \gamma^{\gamma\delta} \I^{\alpha}{}_{\gamma}{}_M \I^{\beta}{}_{\delta}{}_N \varphi^M \varphi^N
\big] \\
&+ \mathcal{O}(\varphi^3)
\end{align*}
for the input coefficients $p^{\alpha\beta}$.
The two remaining sets of input coefficients $E^A{}_\mu$ and $F^A{}_\mu{}^\gamma$ are calculated to take the form
$$\E{\overline{A}}{\mu} = \p{A}[,\mu]\,,\qquad \E{\overline{\overline{A}}}{\mu} = \pp{A}[,\mu]\,,\qquad \E{\overline{\overline{\overline{A}}}}{\mu} = \ppp{A}[,\mu] + \mathcal{O}(\varphi^3)\,,$$
$$\F{\overline{A}}{\mu}{\gamma} = 2\Ii{A}{\mu\sigma}\gu{\sigma\gamma} + 2\Ii{A}{\mu\sigma}\i{\sigma\gamma}{M}\p{M}\,, \qquad \F{\overline{\overline{A}}}{\mu}{\gamma} = -2\Iii{A}{\gamma\sigma}\gd{\sigma\mu} - 2\Iii{A}{\gamma\sigma}\ii{\sigma\mu}{M}\pp{M}\,,$$
$$\F{\overline{\overline{\overline{A}}}}{\mu}{\gamma} = \r{\Iiii{A}{\mu}{\sigma}\iii{\gamma}{\sigma}{M} - \Iiii{A}{\sigma}{\gamma}\iii{\sigma}{\mu}{M}}\ppp{M} + \mathcal{O}(\varphi^3)\,.$$

In the next section, we insert these input coefficients into the gravitational closure equations and evaluate them perturbatively to the order which is required in order to determine linearized gravitational field equations for the spacetime geometry $G$ that underlies weakly birefringent electrodynamics.


\section{Perturbative evaluation of the closure equations}\label{sec_perteval}
Based on the input coefficients that we derived above from weakly birefringent electrodynamics, we are now in a position to set up the closure equations in their perturbative version. 
These are then significantly simplified by curtailing the possible forms of the desired output coefficients such that they yield precisely the terms one requires for linearized gravitational field equations. Taking care of various subtleties concerning the perturbative solution of the closure equations, we finally obtain the weak gravitational field dynamics.  

\subsection{Output coefficients required for linearized theory}\label{sec:linLagrangian}
In order to obtain the linearized gravitational field equations for some given matter, the closure equations on the last two pages of \cite{DSSW} do not need to be solved for all output coefficients $C_{A_1\dots A_N}$, but only for $C$, $C_{A_1}$ and $C_{A_1A_2}$. And even for these three, it suffices to consider, respectively, expansions up to second, first and zeroth order in powers of the configuration variables and their derivatives:
\begin{align}
 C[\varphi] &= \lambda + \sum_{m=0}^{\infty}\Ld{M}{\s1\ldots\s{m}}\P{M}[,\s1\ldots\s{m}] + \frac{1}{2}\sum_{m=0}^{\infty}\sum_{n=0}^{\infty}\LdLd{M}{\s1\ldots\s{m}}{N}{\t1\ldots\t{n}}\P{M}[,\s1\ldots\s{m}]\P{N}[,\t1\ldots\t{n}] + \O3, \label{eq:Cansatz}\\
 C_A[\varphi] &= \xi_{A} + \sum_{m=0}^{\infty}\XXd{A}{M}{\s1\ldots\s{m}}\P{M}[,\s1\ldots\s{m}] + \O2, \label{eq:CAprime} \\
 C_{AB}[\varphi] &= \theta_{AB} + \O1\,,\label{eq:CAB}
\end{align}
where two of the three appearing constant {\it expansion coefficients} $\lambda$, $\xi$ and $\theta$ feature, by construction, the block exchange symmetries
$$
 \LdLd{A}{\mathcal{D}}{B}{\mathcal{E}} = \LdLd{B}{\mathcal{E}}{A}{\mathcal{D}}\qquad\text{and}\qquad 
 \TT{A}{B} = \TT{B}{A},
$$
where $\mathcal{D}$ and $\mathcal{E}$ stand for an arbitrary number of totally symmetric spatial indices, including none at all. This is because any higher orders of these three output coefficients, or indeed any other output coefficients in the first place, do not contribute to the constant and linear order terms of the evolution and constraint equations that are displayed in section IV.E of \cite{DSSW} in terms of the scalar density $\mathcal{L}$, which thus is curtailed to
\begin{equation}
 \mathcal{L}[\varphi,k) = C[\varphi] + C_A[\varphi]k^A + C_{AB}[\varphi]k^Ak^B + \O3\,, \label{eq:Lprime}
\end{equation}
where $\O3$ denotes third and higher powers of either the configuration variables $\varphi^A$, the associated velocities $k^A$ or combinations of both.

A further simplification arises from the fact that no summand in the output coefficient $C_A$ that can be written as the functional gradient $\delta \Lambda/ \delta \varphi^A$ of some scalar functional $\Lambda[\varphi]$ can contribute to the evolution equations that follow from $\mathcal{L}$ according to section IV.E of \cite{DSSW}. We therefore choose to eliminate any such terms from our analysis. To this end, first note that insertion of (\ref{eq:CAprime}) into closure equation (C$18_N$) yields the symmetry conditions 
$$
\XXd{A}{B}{\s1\ldots\s{2n}} = \XXd{B}{A}{\s1\ldots\s{2n}} \qquad\text{and}\qquad \XXd{A}{B}{\s1\ldots\s{2n+1}} = -\XXd{B}{A}{\s1\ldots\s{2n+1}}
$$
for the expansion coefficients $\xi$ of $C_A$. But then letting 
$$\Lambda[\varphi] := \int\text{d}^3y\e{\xi_{M}\P{M}(y) + \frac{1}{2} \xi_{(AB)}\varphi^A(y) \varphi^B(y) + \frac{1}{2}\sum_{n=1}^{\infty}\XXd{M}{N}{\s1\ldots\s{n}}\P{M}(y)\P{N}[,\s1\ldots\s{n}](y)}\,, 
$$
one quickly sees, by direct calculation, that
$$C_A[\varphi] = \XX{[A}{B]}\P{B} + \XXd{(A}{B)}{\alpha}\P{B}[,\alpha] + \frac{\delta\Lambda}{\delta\P{A}}[\varphi] + \O2\,.$$
Since the first two remaining terms cannot be absorbed into the functional gradient, we thus arrive at the final form 
\begin{equation}
C_A[\varphi] = \XX{[A}{B]}\P{B} + \XXd{(A}{B)}{\alpha}\P{B}[,\alpha] + \O2 \label{eq:CA}
\end{equation}
that we will employ in our analysis for this output coefficient. In turn, this dependence of $C_A$ on at most first order derivatives of the configuration variables allows to conclude, by the discussion in section IV.D of \cite{DSSW}, that also $C$ can depend on at most second order derivatives of the $\varphi^A$. Thus we are able to restrict attention to an output coefficient $C$ of the form
\begin{align}
  C[\varphi] &= \lambda + \L{M}\P{M} + \Ld{M}{\sigma}\P{M}[,\sigma] + \Ldd{M}{\s1\s2}\P{M}[,\s1\s2] \nonumber\\
    &\hphantom{=} + \frac{1}{2}\LL{M}{N}\P{M}\P{N} + \LLd{M}{N}{\sigma}\P{M}\P{N}[,\sigma] + \LLdd{M}{N}{\s1\s2}\P{M}\P{N}[,\s1\s2] \nonumber\\
    &\hphantom{=} + \frac{1}{2}\LdLd{M}{\sigma}{N}{\tau}\P{M}[,\sigma]\P{N}[,\tau] + \LdLdd{M}{\sigma}{N}{\t1\t2}\P{M}[,\sigma]\P{N}[,\t1\t2] + \frac{1}{2}\LddLdd{M}{\s1\s2}{N}{\t1\t2}\P{M}[,\s1\s2]\P{N}[,\t1\t2] + \O3\,.\label{eq:C}
\end{align}
Solution of the gravitational closure equations for weakly birefringent electrodynamics thus only requires the determination of the 13 constant tensor expansion coefficients $\theta$, $\xi$ and $\lambda$ that remain in (\ref{eq:CAB}), (\ref{eq:CA}) and (\ref{eq:C}).

\subsection{Closure equations required for linearized theory}\label{sec:evalrules}
We now study the order to which each closure equation must be evaluated. To this end, one first determines, summand by summand and derivative order by derivative order, the non-negative integer
$$\textrm{\it first unknown order of output coefficient} + \sum \text{\it lowest non-zero order of input coefficients}$$
and then considers the minimum of these integers. The result is the lowest order of the equation that cannot be trusted anymore. Thus the equation must be evaluated up to exactly one order less. We demonstrate the procedure for one closure equation, (C$3$) of \cite{DSSW}, which reads
\begin{align*}
0&= 6p^{(\mu|\rho}C_{AB}\F{A}{\rho}{|\nu)} + \sum_{K=0}^{\infty}(K+1)C_{B:A}{}^{\a1\ldots\a{K}(\mu|}\MC{A}{|\nu)}{}_{,\a1\ldots\a{K}}\nonumber \\
    &\hphantom{=} - \sum_{K=0}^{\infty}\r{-1}^K\binom{K+2}{K}\r{\partial^{K}_{\a1\ldots\a{K}}C_{:B}{}^{\a1\ldots\a{K}\mu\nu}}.
\end{align*}
In this closure equation, the first order of $C_{AB}$ is unknown and since $p^{\mu\rho}$ as well as $\F{A}{\rho}{\nu}$ have a zeroth order, their product is unknown in $\O\varphi$. This in turn tells us immediately that this equation can only be evaluated to zeroth order. The first sum contains $M^{A\gamma}$ which has no zeroth order and the second sum is non-zero only for $K=0$ (because $C$ depends only on up to second derivatives of $\varphi$). We thus conclude that $(C$3$)$ must be  evaluated to zeroth order, 
\begin{align*}
 0 &= -6p^{(\mu|\rho}|_{0}C_{AB}\F{A}{\rho}{|\nu)}|_{0} + C_{:B}{}^{\mu\nu}|_{0}\,,
\end{align*}
and also only to zeroth order.

In analogous fashion, one must analyze all closure equations. The remaining set of closure equations, which are required to determine the linearized gravitational field equations that underpin weakly birefringent electrodynamics, are given in the table \ref{table:pde} together with the relevant orders at which they must be evaluated.\\

\begin{longtable}{ccp{0.68\textwidth}}
\hline
	\textsc{Eqn} & \textsc{Required orders} \,\, & \textsc{Remaining pde to be solved to those orders} \\
	\hline
	\endhead
	(C1) & $0$, $1$ & $0 = C\kd{\gamma}{\mu} - 2C_{:A}{}^{\alpha\gamma}\E{A}{\mu,\alpha} + C_{:A}\F{A}{\mu}{\gamma} + C_{:A}{}^{\s1\s2}\F{A}{\mu}{\gamma}{}_{,\s1\s2}$ \\
	(C2) & $0$ & $ 0 = C_{B:A}\F{A}{\mu}{\gamma}$ \\
	(C3) & $0$ & $0 = -6C_{AB}p^{\rho(\mu|}\F{A}{\rho}{|\nu)} + C_{:B}{}^{\mu\nu}$ \\
	(C4) & $0$, $1$ & $0 = 6C_{AB}p^{\mu\nu}\E{A}{\nu} - 6C_{AB}p^{\mu\nu}{}_{,\gamma}\F{A}{\nu}{\gamma} - C_{A}\MC{A}{\mu}{}_{:B} - C_{B:A}\MC{A}{\mu}$ \\
	& & $\hphantom{=} - C_{B:A}{}^{\alpha}\MC{A}{\mu}{}_{,\alpha} - C_{:B}{}^{\mu} + 2C_{:B}{}^{\mu\alpha}{}_{,\alpha}$ \\
	(C5) & $0$, $\varphi\varphi$ & $0 = -6C_{A,\alpha}p^{\sigma\nu}\F{A}{\sigma}{\alpha} + 2C_{:A}\MC{A}{\nu} + C_{:A}{}^{\alpha\beta}\MC{A}{\nu}{}_{,\alpha\beta}$ \\
	(C6) & $0$ & $0 = 4C_{A(M}\MC{A}{\gamma}{}_{:B)} + 2C_{(B:M)}{}^{\gamma}$ \\
	($\text{C8}_{\text{2}}$) & $0$, $1$ & $0 = C_{:A}{}^{\b1\b2} - C_{:A}{}^{(\b1|}\F{A}{\mu}{|\b2)} - 2C_{:A}{}^{\alpha(\b1|}\F{A}{\mu}{|\b2)}{}_{,\alpha}$ \\
	($\text{C8}_{\text{3}}$) & $0$, $1$ & $0 = C_{:A}{}^{(\b1\b2|}\F{A}{\mu}{|\b3)}$ \\
	($\text{C9}_{\text{2}}$) & $0$ & $0 = C_{B:A}{}^{(\b1|}\F{A}{\mu}{|\b2)}$ \\
	($\text{C21}_{\text{3}}$) & $1$, $2$ & $0 = C_{:A}{}^{(\m1\m2|}\MC{A}{|\m3)}$ \\
	\hline
\caption{Closure equations and orders at which they must be evaluated in order to obtain linearized gravitational field equations underpinning weakly birefringent electrodynamics.\label{table:pde}}
\end{longtable}

\subsection{Reduction to a linear homogeneous system of scalar equations}
We will now reduce, in two consecutive steps, the gravitational closure equations above from a system of linear homogeneous partial differential equations with non-constant coefficients into a system of linear equations for a finite set of real numbers, which can then be solved by a standard Gauss-Jordan algorithm. 

To achieve this reduction, the first step is to insert the perturbative forms (\ref{eq:CAB}),(\ref{eq:CA}) and (\ref{eq:C}) of the output coefficients into the partial differential closure equations of table \ref{table:pde} and to thus convert them into a set of 
$$\text{\it eighty linear algebraic relations between the expansion coefficients $\theta$, $\xi$, $\lambda$}$$
listed in the appendix. 

In order to further reduce these equations for the expansion coefficients into a linear system for real scalars, in a second step, we consider the general form which each of the constant expansion coefficients $\lambda$, $\xi$, $\theta$ can take. 
This is a combinatorial exercise that takes care to avoid some,  at first sight maybe unexpected, ambiguities \cite{Wierzba}. For definiteness, we illustrate the procedure for one randomly chosen expansion coefficient, say
$$
\lambda_{\Aa}^{\vphantom{\mu}}{}_{\,\Ba}^{\,\,\mu}\,. 
$$
The key to determining its general form is to note that it can be converted, like any other expansion coefficient, back and forth to a flat tensor 
$$
\lambda_{\overline{\alpha\beta}\vphantom{\overline{\overline{\alpha\beta}}}\,}^{\vphantom{\mu}}{}_{\overline{\gamma\delta}\vphantom{\overline{\overline{\alpha\beta}}}}^{\mu}
$$
on the three-manifold $\Sigma$, by virtue of
$$\lambda_{\overline{\alpha\beta}\vphantom{\overline{\overline{\alpha\beta}}}\,}^{\vphantom{\mu}}{}_{\overline{\gamma\delta}\vphantom{\overline{\overline{\alpha\beta}}}}^{\mu} = \lambda_{\Aa}^{\vphantom{\mu}}{}_{\,\Ba}^{\,\,\mu} \,\, \mathcal{I}^{\Aa}{}_{\alpha\beta} \,\, \mathcal{I}^{\Ba}{}_{\gamma\delta} \qquad \textrm{and} \qquad 
\lambda_{\Aa}^{\vphantom{\mu}}{}_{\,\Ba}^{\,\,\mu} = \lambda_{\overline{\alpha\beta}\vphantom{\overline{\overline{\alpha\beta}}}\,}^{\vphantom{\mu}}{}_{\overline{\gamma\delta}\vphantom{\overline{\overline{\alpha\beta}}}}^{\mu} \,\, \mathcal{I}{}^{\alpha\beta}{}_{\Aa} \,\,\mathcal{I}{}^{\gamma\delta}{}_{\Ba}
\,,$$
which in this case imposes the symmetries $\lambda_{\overline{\alpha\beta}\vphantom{\overline{\overline{\alpha\beta}}}\,}^{\vphantom{\mu}}{}_{\overline{\gamma\delta}\vphantom{\overline{\overline{\alpha\beta}}}}^{\mu} = \lambda_{\overline{(\alpha\beta)}\vphantom{\overline{\overline{\alpha\beta}}}\,}^{\vphantom{\mu}}{}_{\overline{(\gamma\delta)}\vphantom{\overline{\overline{\alpha\beta}}}}^{\mu}$, which will play a role further below.
Now the only available tensorial structures on $\Sigma$, from  which flat tensors could be constructed, are provided by 
$$\gamma_{\alpha\beta}\,,\quad \gamma^{\alpha\beta}\,,\quad\epsilon_{\alpha\beta\gamma}\,,\quad \delta^\alpha_\beta$$
and arbitrary products and contractions thereof. So in order to determine the general form of $\lambda_{\overline{\alpha\beta}\vphantom{\overline{\overline{\alpha\beta}}}\,}^{\vphantom{\mu}}{}_{\overline{\gamma\delta}\vphantom{\overline{\overline{\alpha\beta}}}}^{\mu}$, we first write down all possible monomial terms that can be constructed from those and possess the correct index structure. For our example, these are the 10 monomials
$$ \epsilon{}_{\alpha\beta\gamma}\delta{}_{\delta}^{\mu}\,, \quad
    \epsilon{}_{\alpha\beta\delta}\delta{}_{\gamma}^{\mu}\,, \quad
    \epsilon{}_{\alpha\beta}{}^{\mu}\gamma{}_{\gamma\delta} \,,\quad
    \epsilon{}_{\alpha\gamma\delta}\delta{}_{\beta}^{\mu}\,,\quad
    \epsilon{}_{\alpha\gamma}{}^{\mu}\gamma{}_{\beta\delta} \,,
$$
$$
    \epsilon{}_{\alpha\delta}{}^{\mu}\gamma{}_{\beta\gamma} \,, \quad
    \epsilon{}_{\beta\gamma\delta}\delta{}_{\alpha}^{\mu} \,,
    \quad
    \epsilon{}_{\beta\gamma}{}^{\mu}\gamma{}_{\alpha\delta} \,, \quad
    \epsilon{}_{\beta\delta}{}^{\mu}\gamma{}_{\alpha\gamma} \,,\quad
     \epsilon{}_{\gamma\delta}{}^{\mu}\gamma{}_{\alpha\beta}\,,
$$
where further, obviously linearly dependent monomials, such as $\epsilon{}_{\beta\alpha\gamma}\delta{}_{\delta}^{\mu}$, have been omitted. These omissions, however, are just economical but by no means essential, since any linear dependence would be automatically eliminated in the following step. For there is a subtlety, due to the existence of a chart where the components of the background tensors only take on values $-1$, $0$ and $1$, which causes some additional linear dependencies that go unnoticed when only considering the abstract tensor structure. In order to capture these hidden linear dependencies, as well as any of those which one might have failed to spot when providing the initial list of all possible monomials with the correct index structure, one proceeds as follows. For each of the tensor monomials listed above, one assigns a $3^\text{rank}$-tuple to each tensor monomial listed above, collecting the  values of all of its components, which for our example amounts to constructing the $3^5$-tuples  
$$\epsilon{}_{\alpha\beta\gamma}\delta{}_{\delta}^{\mu} \mapsto \left(
	\begin{array}{c}
		\epsilon_{111}\delta^1_1 \\
		\epsilon_{111}\delta^1_2 \\
		\epsilon_{111}\delta^1_3 \\
		\epsilon_{111}\delta^2_1 \\
		\epsilon_{111}\delta^2_2 \\
		\epsilon_{111}\delta^2_3 \\
		\dots
	\end{array}
\right) \,, \quad \epsilon{}_{\alpha\beta\delta}\delta{}_{\gamma}^{\mu} \mapsto \left( \begin{array}{c}
		\epsilon_{111}\delta^1_1 \\
		\epsilon_{112}\delta^1_1 \\
		\epsilon_{113}\delta^1_1 \\
		\epsilon_{111}\delta^2_1 \\
		\epsilon_{112}\delta^2_1 \\
		\epsilon_{113}\delta^2_1 \\
		\dots
	\end{array}
\right) \,, \quad \epsilon{}_{\alpha\beta}{}^{\mu}\gamma{}_{\gamma\delta} \mapsto \left( \begin{array}{c}
		\epsilon_{11}{}^{1}\gamma_{11} \\
		\epsilon_{11}{}^{1}\gamma_{12} \\
		\epsilon_{11}{}^{1}\gamma_{13} \\
		\epsilon_{11}{}^{2}\gamma_{11} \\
		\epsilon_{11}{}^{2}\gamma_{12} \\
		\epsilon_{11}{}^{2}\gamma_{13} \\
		\dots
	\end{array}
\right) \,, \quad \dots \,,
$$
where $\alpha\beta\gamma\mu\delta$ is taken to run in the same order, say the lexicographic one employed above, for each tuple. In order to identify linearly dependent monomials, one now  performs a Gauss algorithm on the matrix whose columns are provided by the above tuples. Taking the corresponding $3^5 \times 10$ matrix of our example to row echelon form reveals that the last four monomials can indeed be expressed as linear combinations of the first six monomials, namely
\begin{eqnarray*}
\epsilon{}_{\beta\gamma\delta}\delta{}_{\alpha}^{\mu} = \epsilon{}_{\alpha\beta\gamma}\delta{}_{\delta}^{\mu} - \epsilon{}_{\alpha\beta\delta}\delta{}_{\gamma}^{\mu} + \epsilon{}_{\alpha\gamma\delta}\delta{}_{\beta}^{\mu} \,, \quad&\quad&\quad 
    \epsilon{}_{\beta\gamma}{}^{\mu}\gamma{}_{\alpha\delta} =  \epsilon{}_{\alpha\beta\gamma}\delta{}_{\delta}^{\mu} - \epsilon{}_{\alpha\beta}{}^{\mu}\gamma{}_{\gamma\delta} + \epsilon{}_{\alpha\gamma}{}^{\mu}\gamma{}_{\beta\delta} \,, \\
    \epsilon{}_{\beta\delta}{}^{\mu}\gamma{}_{\alpha\gamma} = \epsilon{}_{\alpha\beta\delta}\delta{}_{\gamma}^{\mu} - \epsilon{}_{\alpha\beta}{}^{\mu}\gamma{}_{\gamma\delta} + \epsilon{}_{\alpha\delta}{}^{\mu}\gamma{}_{\beta\gamma} \,, \quad&\quad&\quad
     \epsilon{}_{\gamma\delta}{}^{\mu}\gamma{}_{\alpha\beta} = \epsilon{}_{\alpha\gamma\delta}\delta{}_{\beta}^{\mu} - \epsilon{}_{\alpha\gamma}{}^{\mu}\gamma{}_{\beta\delta} + \epsilon{}_{\alpha\delta}{}^{\mu}\gamma{}_{\beta\gamma}\,.
\end{eqnarray*}
This leaves us with the first six of the ten monomials we initially listed. These would provide the basis from which one could generate the general form of the flat tensor in our example by taking an arbitary linear combination with undetermined scalar coefficients. But before doing so, one needs to impose on each of these remaining six monomials all symmetries that the
$\lambda_{\overline{\alpha\beta}\vphantom{\overline{\overline{\alpha\beta}}}\,}^{\vphantom{\mu}}{}_{\overline{\gamma\delta}\vphantom{\overline{\overline{\alpha\beta}}}}^{\mu}$ inherited from its construction. Indeed, symmetrization in $\alpha\beta$ further reduces the basis to merely  three monomials
$$
    \epsilon{}_{(\alpha|\gamma\delta}\delta{}_{|\beta)}^{\mu} \,, \qquad 
    \epsilon{}_{(\alpha|\gamma}{}^{\mu}\gamma{}_{|\beta)\delta}\,, \qquad
    \epsilon{}_{(\alpha|\delta}{}^{\mu}\gamma{}_{|\beta)\gamma}\,,	
$$
which, by symmetrization in $\gamma\delta$, again is narrowed down to only one monomial, namely $\epsilon{}_{(\alpha|(\gamma}{}^{\mu}\gamma{}_{\delta)|\beta)}$, so that the general form of the expansion coefficient we studied is
$$
\lambda_{\Aa}^{\vphantom{\mu}}{}_{\,\Ba}^{\,\,\mu} =  c_1 \cdot  \epsilon{}_{(\alpha|(\gamma}{}^{\mu}\gamma{}_{\delta)|\beta)} \,\, \mathcal{I}{}^{\alpha\beta}{}_{\Aa} \,\,\mathcal{I}{}^{\gamma\delta}{}_{\Ba}
$$
for some constant $c_1$. 
Precisely along the same lines, one determines the general form for all expansion coefficients $\theta$, $\xi$ and $\lambda$.

This systematic procedure for generating the general form of the expansion coefficients has been cast into the form of a computer program \footnote{Publicly available at \url{http://www.github.com/constructivegravity/construct}}, which has been used in the next subsection to solve the linear system of equations one obtains after insertion of the general forms of the expansion coefficients into the pertinent closure equations listed in table \ref{table:pde}.

\subsection{Gravitational Lagrangian}\label{sec_Lagrangian}
All that remains, in order to determine the Lagrangian that yields the linearized gravitational field equations for the geometry underpinning weakly birefringent electrodynamics, is to solve the linear homogeneous system for the constant scalars that emerges when the general forms for the expansion coefficients $\theta$, $\xi$ and $\lambda$ are inserted into  the eighty relations listed in the appendix; see also \cite{Nthesis,SchneiderMSc}. Using the shorthand notations
$$
\dofa{\alpha\beta} := \I^{\alpha\beta}{}_{\Aa}\,\varphi^{\Aa}\,,\qquad \dofb{\alpha\beta} := \I_{\alpha\beta\,}{}_{\Ab}\,\varphi^{\Ab}\,,\qquad \dofc{\alpha}{\beta} := \I^{\alpha}{}_{\beta}{}_{\Ac}\, \varphi^{\Ac}\,,
$$
and similar notations for the generalized velocities $k^A$, one obtains the gravitational Lagrangian in terms of  
$$
\textrm{\it twenty-two undetermined constants }\, g_1, g_2, \dots, g_{22}\,
$$ 
only, which is of course a drastic reduction compared to the 136 undetermined constants in the general forms before evaluation of the closure equations, which any speculative treatment without closure equations would have had to determine experimentally.   
But then again only eleven independent combinations of these twenty-two constants turn out to survive in the associated gravitational equations of motion, as will be discussed later. We abstained from using this hindsight at this stage in order to not obscure the derivation. The complete result for the scalar density $\mathcal{L}$ is displayed overleaf. 

Note that $\mathcal{L}$ is not quite yet the actual gravitational Lagrangian $\mathscr{L}_{\tiny geometry}$ in terms of spacetime fields. The latter, however, is readily constructed from the former, as shown in section IV.E of \cite{DSSW}. There it is also shown how, alternatively, the gravitational equations of motion can be written down directly in terms of the scalar density $\mathcal{L}$ whose expansion (\ref{eq:Lprime}) emerges immediately from the solution of the gravitational closure equations. 
\newpage
$ $
\vspace{-.85cm}
\small{
\begin{flalign*}
&\mathcal{L}(\varphi,\partial \varphi, \partial\partial \varphi, k) = 
	4 \left(g_{18}-2 g_{19}+g_{20}\right) + 
	\Big[2\left(g_{19}-g_{18}\right) \gamma_{\alpha\beta}\Big] \dofa{\alpha\beta} + 
	\Big[2\left(g_{20} - g_{19}\right) \gamma^{\alpha\beta}\Big] \dofb{\alpha\beta} \\& + \Big[\tfrac{1}{3} (g_4-4 g_2
	-g_3+g_6-2 g_7) \,\gamma_{\alpha\beta}\gamma^{\mu\nu} + (g_3+g_6-6 g_7-2g_9)\, \delta_{\alpha}^{\mu} \delta_{\beta}^{\nu} \Big]  \dofa[\mu\nu]{\alpha\beta}\\&
    + \Big[\tfrac{1}{3} (8 g_3-4 g_2+g_4+4 g_6 -44 g_7+12 g_8-12 g_9)\,\gamma^{\alpha\beta}\gamma^{\mu\nu} + 
	2 (4 g_7-g_3-2 g_8+g_9) \, \gamma^{\mu\alpha}\gamma^{\beta\nu} \Big] \dofb[\mu\nu]{\alpha\beta} \\&
    + \Big[g_{18} \, \gamma_{\alpha\beta} \gamma_{\mu\nu} + (2 g_{18}-4 g_{19}+2 g_{20}+g_{21}) \, \gamma_{\alpha\mu}\gamma_{\nu\beta}\Big] \dofa{\alpha\beta}\dofa{\mu\nu} + 
	\Big[g_{20} \gamma^{\alpha\beta} \gamma^{\mu\nu} + g_{21} \gamma^{\alpha\mu}\gamma^{\nu\beta} \Big] \dofb{\alpha\beta} \dofb{\mu\nu} \\&
    + \Big[g_{19} \gamma_{\alpha\beta} \gamma^{\mu\nu} + (2 g_{20}-2 g_{19}+g_{21}) \delta_{\alpha}^\mu \delta^\nu_\beta \Big] \dofa{\alpha\beta} \dofb{\mu\nu} + 
	\Big[g_{22} \gamma_{\mu\alpha} \delta^\nu_\beta\Big] \dofa{\alpha\beta} \dofc{\mu}{\nu} + 
	\Big[g_{22} \gamma^{\nu\alpha}\delta^\beta_\mu\Big]  \dofb{\alpha\beta} \dofc{\mu}{\nu} \\&
    + \Big[4 (g_{19}-g_{20}-g_{21})\Big] \dofc{\alpha}{\beta} \dofc{\beta}{\alpha}
    + \Big[ g_{17}\, \epsilon_{\alpha\mu}{}^\lambda \gamma_{\beta\nu} \Big] \dofa{\alpha\beta} \dofa[\lambda]{\mu\nu} + 
	\Big[ g_{17}\, \epsilon_{\alpha}{}^{\mu\lambda} \delta_{\beta}^\nu\Big] \dofa{\alpha\beta} \dofb[\lambda]{\mu\nu}\\&
    +\Big[g_{17}\, \epsilon^{\alpha}{}_{\mu}{}^{\lambda} \delta^{\beta}_\nu\Big] \dofb{\alpha\beta} \dofa[\lambda]{\mu\nu}
    + \Big[g_{17}\, \epsilon^{\alpha\mu\lambda} \gamma^{\beta\nu}\Big] \dofb{\alpha\beta} \dofb[\lambda]{\mu\nu} + 
	\Big[4g_{17} \epsilon_{\alpha\mu}{}^{\lambda} \gamma^{\beta\nu}\Big] \dofc{\alpha}{\beta} \dofc[\lambda]{\mu}{\nu} \\&
    + \Big[\tfrac{1}{6} (6 g_1+4 g_2+g_3-g_4-g_6+2 g_7) \gamma_{\alpha\beta}\gamma_{\mu\nu}\gamma^{\lambda\kappa}  + 
	\tfrac{1}{3} (4 g_2+g_3-g_4-g_6+2 g_7+3 g_{10}) \gamma_{\alpha\mu}\gamma_{\nu\beta} \gamma^{\lambda\kappa} \\&
    \quad\,\, + \tfrac{1}{4} (g_4+2 g_6-8 g_7-2 g_9) \gamma_{\alpha\beta}\delta^\lambda_\mu \delta_\nu^\kappa +
	g_3 \, \gamma_{\alpha\mu} \delta^\lambda_\nu \delta^\kappa_\beta +
	\tfrac{1}{3} (g_4-g_2-g_3+g_6-2 g_7) \gamma_{\mu\nu} \delta^\lambda_\alpha \delta^\kappa_\beta	\Big] \dofa{\alpha\beta} \dofa[\lambda\kappa]{\mu\nu} \\&
    +\Big[\tfrac{1}{2} (2 g_1+g_3+g_6-6 g_7-2 g_9) \gamma_{\alpha\beta} \gamma^{\mu\nu} \gamma^{\lambda\kappa} + \tfrac{1}{12} (8 g_2-4 g_3+g_4-2 g_6+16 g_7+6 g_9) \gamma_{\alpha\beta}\gamma^{\mu\lambda}\gamma^{\kappa\nu}\\& 
    \quad\,\, + \tfrac{1}{2} (g_3-g_6-2 g_7+4 g_8+2 g_{10}) \delta_\alpha^\mu \delta_\beta^\nu \gamma^{\lambda\kappa} + 
	\tfrac{1}{6} (7 g_3-2 g_2+2 g_4+5 g_6-46 g_7+12 g_8-12 g_9) \gamma^{\mu\nu} \delta^\lambda_\alpha \delta^\kappa_\beta \\&
    \quad\,\,+ (g_6+2 g_7-4 g_8 ) \delta_\alpha^\mu \gamma^{\nu\lambda}\delta^\kappa_\beta \Big] \dofa{\alpha\beta} \dofb[\lambda\kappa]{\mu\nu} \\&
    + \Big[\tfrac{1}{2}g_{14} \, \gamma_{\alpha\beta} \gamma^{\nu\lambda} \delta^{\kappa}_\mu + \tfrac{1}{2}(2 g_{12}+g_{13}) \gamma_{\alpha\mu} \gamma^{\lambda\kappa} \delta_{\beta}^{\nu}
    \,\,+ g_{15} \, \gamma_{\alpha\mu} \gamma^{\nu\lambda} \delta^{\kappa}_{\beta} \Big] \dofa{\alpha\beta} \dofc[\lambda\kappa]{\mu}{\nu}\\&
	+\Big[
	g_1 \, \gamma^{\alpha\beta}\gamma_{\mu\nu} \gamma^{\lambda\kappa} +
	\tfrac{1}{4} (2 g_3+g_4+4 g_6-20 g_7-6 g_9) \gamma^{\alpha\beta} \delta^\lambda_\mu \delta^\kappa_\nu 
    \,\, + g_{10} \, \delta^\alpha_\mu \delta^\beta_\nu \gamma^{\lambda\kappa} +
	g_3 \, \gamma^{\alpha\lambda} \delta^\kappa_\mu \delta^\beta_\nu \\&
    \quad\,\,+ 
	g_2 \, \gamma^{\alpha\lambda}\gamma_{\mu\nu} \gamma^{\kappa\beta}
	\Big]\dofb{\alpha\beta} \dofa[\lambda\kappa]{\mu\nu} \\&
	+\Big[\tfrac{1}{6} (6 g_1-4 g_2+11 g_3+g_4+7 g_6-62 g_7+12 g_8-18 g_9) \gamma^{\alpha\beta} \gamma^{\mu\nu} \gamma^{\lambda\kappa}\\& 
    \quad\,\,+ \tfrac{1}{12} (8 g_2-16 g_3+g_4-2 g_6+64 g_7-24 g_8+18 g_9) \gamma^{\alpha\beta}\gamma^{\mu\lambda}\gamma^{\kappa\nu}\\& 
 \quad\,\,+ \tfrac{1}{6} (8 g_2-13 g_3-2 g_4-11 g_6+82 g_7-12 g_8+24 g_9+6 
g_{10}) \gamma^{\lambda\kappa} \gamma^{\alpha\mu}\gamma^{\nu\beta} + (4 g_3+g_6-14 g_7\\&
\quad\,\, +4 g_8-4 g_9) \gamma^{\lambda\alpha}\gamma^{\beta\mu}\gamma^{\nu\kappa} + 
\tfrac{1}{2}( 2 g_2- 3 g_3-g_6+14 g_7-4 g_8+4 g_9 ) \gamma^{\mu\nu} \gamma^{\alpha\lambda}\gamma^{\kappa\beta}
	\Big] \dofb{\alpha\beta} \dofb[\lambda\kappa]{\mu\nu} \\&
    + \Big[\tfrac{1}{2}g_{14} \, \gamma^{\alpha\beta}\gamma^{\nu\lambda}\delta^\kappa_\mu + \tfrac{1}{2}(2 g_{12}+g_{13}) \gamma^{\lambda\kappa} \gamma^{\nu\alpha} \delta^\beta_\mu + g_{15} \, \gamma^{\alpha\lambda}\gamma^{\kappa\nu}\delta^{\beta}_\mu  \Big] \dofb{\alpha\beta} \dofc[\lambda\kappa]{\mu}{\nu} \\&
    + \Big[\tfrac{1}{2}(2g_{12}+g_{13}) \gamma_{\alpha\mu} \gamma^{\lambda\kappa} \delta^\beta_\nu + g_{15} \, \gamma_{\alpha\mu}\gamma^{\beta\lambda} \delta_\nu^\kappa + \tfrac{1}{2}g_{14} \, \gamma_{\mu\nu} \gamma^{\beta\lambda} \delta_\alpha^\kappa\Big]
	\dofc{\alpha}{\beta} \dofa[\lambda\kappa]{\mu\nu} \\&
    + \Big[\tfrac{1}{2}(2g_{12}+g_{13}) \gamma^{\beta\mu} \gamma^{\lambda\kappa} \delta_\alpha^\nu + g_{15} \, \gamma^{\beta\lambda} \gamma^{\kappa\mu} \delta^\nu_\alpha + \tfrac{1}{2}g_{14} \, \gamma^{\mu\nu} \gamma^{\beta\lambda} \delta_\alpha^\kappa\Big]\dofc{\alpha}{\beta} \dofb[\lambda\kappa]{\mu\nu}
    \\
     &  +\Big[\frac{1}{3} (13 g_3-8 g_2+2 g_4+11 g_6-82 g_7+12 g_8-24 g_9-12 g_{10}) \gamma_{\alpha\mu}\gamma^{\beta\nu}\gamma^{\lambda\kappa}\\& 
    \quad\,\,-2 (3 g_3+g_6-6 g_7-2g_9) \gamma_{\alpha\mu} \gamma^{\beta\lambda} \gamma^{\kappa\nu} \Big] \dofc{\alpha}{\beta} \dofc[\lambda\kappa]{\mu}{\nu} 
    \\
	& + \dots
\end{flalign*}}
$ $
\vspace{-.85cm}
\small{
\begin{flalign*}
      & + \Big[\tfrac{1}{2} g_4 \left(\gamma_{\alpha\beta}\delta^\lambda_\mu \delta^\kappa_\nu + \gamma_{\mu\nu}\delta^\lambda_\alpha \delta^\kappa_\beta \right) + 
		g_5 \gamma_{\alpha\beta} \gamma_{\mu\nu} \gamma^{\lambda\kappa} + 
		g_6 \gamma_{\alpha\mu} \delta_\beta^\lambda \delta_\nu^\kappa + 
		g_{11} \gamma_{\alpha\mu}\gamma_{\nu\beta} \gamma^{\lambda\kappa} +
		g_9 \gamma_{\alpha\mu} \delta_\nu^\lambda \delta_\beta^\kappa
	\Big]
	\dofa[\lambda]{\alpha\beta} \dofa[\kappa]{\mu\nu}
	\\
    & + \Big[\tfrac{1}{6} (g_4-4 g_2-g_3+6 g_5+g_6-2 g_7)
    \gamma_{\alpha\beta}\gamma^{\mu\nu}\gamma^{\lambda\kappa} + \tfrac{1}{6} (8 g_2+2 g_3+g_4-2 g_6+4 g_7)
    \gamma_{\alpha\beta}\gamma^{\mu\lambda}\gamma^{\kappa\nu} \\& \quad\,\,+
    \tfrac{1}{2}(2 g_3+g_4+2 g_6-12 g_7-4 g_9) \gamma^{\mu\nu} \delta^\lambda_\alpha \delta^\kappa_\beta +
    (g_3+g_6-6 g_7-g_9) \gamma^{\mu\lambda} \delta^\kappa_\alpha \delta^\nu_\beta \\& \quad\,\,+ 
    \tfrac{1}{6} (4 g_4-16 g_2-7 g_3+g_6+10 g_7+6 g_9+6 g_{11})
    \gamma^{\lambda\kappa}\delta^\mu_\alpha \delta^\nu_\beta +
    g_6 \,\gamma^{\mu\kappa} \delta^\lambda_\alpha \delta^\nu_\beta   
    \Big] \dofa[\lambda]{\alpha\beta} \dofb[\kappa]{\mu\nu} \\&
	+ \Big[g_{14} \, \gamma_{\alpha\beta} \gamma^{\nu\lambda}\delta^\kappa_\mu + 2 g_{15} \, \gamma_{\alpha\mu}\gamma^{\nu(\lambda}\delta_\beta^{\kappa)} + g_{13} \gamma_{\alpha\mu} \gamma^{\lambda\kappa} \delta_\beta^\nu\Big] \dofa[\lambda]{\alpha\beta} \dofc[\kappa]{\mu}{\nu} \\&
	+ \Big[\tfrac{1}{6} (8 g_2-10 g_3+g_4-2 g_6
	+52 g_7-24 g_8+12 g_9) \left(\gamma^{\alpha\beta} \gamma^{\lambda\mu}\gamma^{\nu\kappa} + \gamma^{\mu\nu} \gamma^{\lambda\alpha}\gamma^{\beta\kappa}\right) \\&
    \quad\,\,+ \tfrac{1}{6} (7 g_3-8 g_2+2 g_4+6 g_5+5 g_6-46 g_7+12 g_8-12 g_9\
) \gamma^{\alpha\beta} \gamma^{\mu\nu}\gamma^{\lambda\kappa}\\& 
\quad\,\,+(4 g_3+g_6-16 g_7+8 g_8-4 g_9) \gamma^{\lambda\alpha}\gamma^{\beta\mu}\gamma^{\nu\kappa} + \tfrac{1}{2}(54 g_7-11 g_3-5 g_6-12 g_8+16 g_9+2 g_{11}) \gamma^{\alpha\mu}\gamma^{\nu\beta}\gamma^{\kappa\lambda} \\&
    \quad\,\,+ (3 g_3+g_6-14 g_7+4 g_8-3 g_9) \gamma^{\lambda\mu}\gamma^{\nu\alpha}\gamma^{\beta\kappa}
	\Big] \dofb[\lambda]{\alpha\beta} \dofb[\kappa]{\mu\nu} \\& 
    + \Big[g_{14} \gamma^{\alpha\beta}\gamma^{\nu(\kappa}\delta^{\lambda)}_\mu + 2g_{15}\gamma^{\alpha(\kappa}\gamma^{\lambda)\nu}\delta^\beta_\mu + g_{13}\, \gamma^{\lambda\kappa}\gamma^{\alpha\nu}\delta^\beta_\mu 
	\Big] \dofb[\lambda]{\alpha\beta} \dofc[\kappa]{\mu}{\nu}\\&
    + \Big[
	(4 g_8-7 g_3-g_6-2 g_7+4 g_9) \gamma^{\nu\kappa}\delta^\lambda_\alpha\delta^\beta_\mu + \frac{2}{3} (8 g_2+26 g_3-2 g_4+g_6-56 g_7-27 g_9-6 g_{11}) \gamma_{\alpha\mu} \gamma^{\beta\nu} \gamma^{\lambda\kappa}\\&
    \quad\,\,+ 4 (2 g_7-2 g_3+2 g_8+g_9) \gamma_{\alpha\mu} \gamma^{\beta\kappa}\gamma^{\lambda\nu}
	\Big] \dofc[\lambda]{\alpha}{\beta} \dofc[\kappa]{\mu}{\nu}
	\\& 
    + \Big[g_{16} \, \gamma_{\alpha\mu} \delta_\beta^\nu\Big] \vela{\alpha\beta} \dofc{\mu}{\nu} + 
	\Big[g_{16} \, \gamma^{\alpha\nu} \delta^\beta_\mu\Big] \velb{\alpha\beta} \dofc{\mu}{\nu} +
	\Big[(g_{16}+g_{17}) \gamma_{\mu\alpha}\delta^\beta_\nu\Big] \velc{\alpha}{\beta} \dofa{\mu\nu} +
	\Big[(g_{16}+g_{17}) \gamma^{\mu\beta}\delta^\nu_\alpha\Big] \velc{\alpha}{\beta} \dofb{\mu\nu} \\&
    +\Big[(g_6-g_3+2 g_7) \epsilon_{\alpha\mu}{}^\lambda \delta_\beta^\nu\Big] \vela{\alpha\beta} \dofc[\lambda]{\mu}{\nu} 
    + \Big[4 g_8 \, \epsilon^{\alpha}{}_{\mu}{}^\lambda \gamma^{\beta\nu}\Big] \velb{\alpha\beta} \dofc[\lambda]{\mu}{\nu} +
	\Big[4 g_7 \, \epsilon_{\alpha\mu}{}^\lambda \delta^\beta_\nu\Big] \velc{\alpha}{\beta} \dofa[\lambda]{\mu\nu} \\&
    +\Big[4 g_7 \, \epsilon_{\alpha}{}^{\mu\lambda} \gamma^{\beta\nu}\Big]\velc{\alpha}{\beta} \dofb[\lambda]{\mu\nu} + \Big[g_{12} \, \gamma_{\alpha\mu} \delta^\nu_\beta\Big] \vela{\alpha\beta} \velc{\mu}{\nu} + \Big[g_{12} \, \gamma^{\alpha\nu} \delta^\beta_\mu\Big] \velb{\alpha\beta} \velc{\mu}{\nu}\\& 
    +
	\left[
	\tfrac{1}{4}(4 g_1+ 4 g_2-g_4-2g_5)\gamma_{\alpha\beta}\gamma_{\mu\nu} + 
	\tfrac{1}{6} (8 g_2+5 g_3-2 g_4-5 g_6+10 g_7+6 g_{10}-3 g_{11})\gamma_{\alpha\mu}\gamma_{\nu\beta}
	\right] \vela{\alpha\beta} \vela{\mu\nu}\\&
    + \Big[ 
	\tfrac{1}{12} (12 g_1+8 g_2-g_3-2 g_4-6 g_5+g_6-2 g_7) \gamma_{\alpha\beta} \gamma^{\mu\nu} + 
	\tfrac{1}{12} (16 g_2+13 g_3-4 g_4-7 g_6+2 g_7-6 g_9\\&
    \quad\,\,+12 g_{10}-6 g_{11}) \delta_\alpha^\mu \delta_\beta^\nu 
	\Big] \vela{\alpha\beta} \velb{\mu\nu} 
	 +
	 \Big[\tfrac{1}{3} (4 g_4-16 g_2-10 g_3+7 g_6-8 g_7+3 g_9-12 g_{10}+6 g_{11})\Big] \velc{\alpha}{\beta} \velc{\beta}{\alpha}\\&
     + \Big[
	\tfrac{1}{12} (12 g_1+4 g_2+7 g_3-g_4-6 g_5+5 g_6-46 g_7+12 g_8-12 g_9) \gamma^{\alpha\beta}\gamma^{\mu\nu} + 
	\tfrac{1}{12} (16 g_2+7 g_3-4 g_4 \\&
    \quad\,\, -7 g_6+26 g_7-12 g_8+12 g_{10}-6 g_{11}) \gamma^{\alpha\mu}\gamma^{\nu\beta}
	\Big] \velb{\alpha\beta} \velb{\mu\nu} \\& + 
	\mathcal{O}(3)
\end{flalign*}}
The complexity of this result attests the usefulness of the gravitational closure mechanism. It would have been hard to guess the intricate interdependency of the prefactors appearing in this scalar density $\mathcal{L}$, and hence in the gravitational Lagrangian $\mathscr{L}_{\tiny geometry}$, in terms of the independent constants $g_1,  \dots, g_{22}$, or indeed the 11 independent linear combinations $\kappa_1, \dots, \kappa_{11}$ that remain in the ultimately relevant gravitational equations of motion discussed in the following section.

\section{Linearized gravitational equations of motion}\label{sec:gaugedeom}

In this section we present the linearized gravitational field equations for the geometry that underpins birefringent electrodynamics. Due to the diffeomorphism invariance of the theory, there is a gauge ambiguity that must be fixed in order to render the field equations no longer underdetermined. We identify the gauge transformations for the geometry at hand and find a complete set of gauge invariant quantities. Employing a particular gauge, we can then display the decoupled scalar, vector and tensor equations of motion.

\subsection{Gauge transformations and gauge invariants in linearized theory}
It is straightforward to uniquely decompose the 17 perturbative configuration variables $\varphi^A$ according to the well-known Helmholtz decomposition in three spatial dimensions as
\begin{eqnarray*}
\varphi^{\Aa} &=& \I^{\Aa}{}_{\alpha\beta} \Big[ \, \widetilde{F} \,\gamma^{\alpha\beta} +  \Delta^{\alpha\beta} F + 2 \partial^{(\alpha} F^{\beta)} + F^{\alpha\beta}    \Big] \,,\\
\varphi^{\Ab} &=& \I^{\Ab}{}^{\alpha\beta} \Big[\, \widetilde{E}\, \gamma_{\alpha\beta} + \Delta^{\alpha\beta} E + 2 \partial^{(\alpha} E^{\beta)} + E^{\alpha\beta}    \Big] \,,\\
\varphi^{\Ac} &=& \I^{\Ac}{}_{\alpha}{}^{\beta} \Big[ \Delta^\alpha{}_\beta C + \partial^\alpha C_\beta + \partial_\beta C^\alpha + C^\alpha{}_\beta \Big] \, + \mathcal{O}(2)\,,
\end{eqnarray*}
where $\widetilde F, \widetilde E, F, E, C$ are scalar fields, $F^\alpha, E^\alpha, C^\alpha$ are solenoidal vector fields and $F^{\alpha\beta}, E^{\alpha\beta}, C^{\alpha\beta}$ are transverse and $\gamma$-traceless symmetric tensor fields on three-dimensional Euclidean space. The trace-removed Hesse operator $\Delta^{\alpha\beta} := \partial^\alpha \partial^\beta - \tfrac{1}{3}\gamma^{\alpha\beta} \Delta$ and indices are raised and lowered with the Euclidean metric $\gamma$. 
Including Helmholtz-decomposed lapse $N=1+A$ and shift $N^\alpha = \partial^\alpha B + B^\alpha$ fields, the components of the perturbation $H^{abcd}$ of the spacetime geometry thus become
\begin{eqnarray}
H^{0\beta0\delta} &=& (2A-\widetilde F) \gamma^{\beta\delta} - \Delta^{\beta\delta} F - 2 \partial^{(\beta} F^{\delta)} - F^{\beta\delta} \,,\nonumber\\
H^{0\beta\gamma\delta} &=& \epsilon^{\gamma\delta}{}_\nu \Big[(\tfrac{3}{2} \widetilde F - A) \gamma^{\beta\nu} + \epsilon^{\beta\nu\rho}(\partial_\rho B + B_\rho)  + \Delta^{\beta\nu} C + 2 \partial^{(\beta} C^{\nu)} + C^{\beta\nu} \Big]\,,\label{HelmH}\\
H^{\alpha\beta\gamma\delta} &=& \epsilon^{\lambda\alpha\beta} \epsilon^{\kappa\gamma\delta} \Big[(3 \widetilde F + \widetilde E) \gamma_{\lambda\kappa} + \Delta_{\lambda\kappa} E + 2 \partial_{(\lambda} E_{\kappa)} + E_{\lambda\kappa}\Big]\,.\nonumber
\end{eqnarray}

Under a change of gauge that is parameterized by a smooth vector field $\xi$,
the components of the perturbation tensor field $H$ pick up the changes 
\begin{equation*}
\left(\Delta_{\xi} H \right)^{abcd} := \left(\mathcal{L}_{\xi}N \right)^{abcd}\,,
\label{gauge_invariants_change_under_transformation}
\end{equation*}
where $N^{abcd} := \eta^{ac}\eta^{bd} - \eta^{ad}\eta^{bc} - \epsilon^{abcd}$ constitutes the flat area metric background. Decomposing $\xi^0=:T$ and $\xi^\alpha =: \partial^\alpha L + L^\alpha$ in terms of scalar fields $T, L$ and a solenoidal vector field $L^\alpha$, one obtains the explicit form 
\begin{eqnarray}
\left(\Delta_{\xi} H \right)^{0\beta 0\delta} &=& 2 \partial^\beta\partial^\delta L + 2 \partial^{(\beta} L^{\delta)} + 2 \dot T \gamma^{\beta\delta}\,,\nonumber
\\
\left(\Delta_{\xi} H \right)^{0\beta \gamma\delta} &=& \epsilon^{\gamma\delta}{}_\nu \Big[ (- \dot T - \Delta L) \gamma^{\beta\nu} + ( \dot L_\rho + \partial_\rho \dot L - \partial_\rho T) \epsilon^{\beta\nu\rho}\Big]\,,\label{DeltaH}
\\
\left(\Delta_{\xi} H \right)^{\alpha\beta\gamma\delta} &=& \epsilon^{\alpha\beta\kappa} \epsilon^{\gamma\delta\lambda} \Big[2 \partial_\kappa\partial_\lambda L +  2\partial_{(\kappa} L_{\lambda)} - 2 \gamma_{\kappa\lambda} \Delta L  \Big] \,.\nonumber
\end{eqnarray}
The changes (\ref{DeltaH}) in the components of $H$ are induced from corresponding changes 
$$\Delta_\xi \widetilde F = - \tfrac{2}{3} \Delta L\,,\qquad \Delta_\xi A = \dot T \,,\qquad \Delta_\xi \widetilde E = \tfrac{2}{3} \Delta L \,\qquad \Delta_\xi B = \dot L - T\,,$$
$$\Delta_\xi F = - 2L\,,\qquad \Delta_\xi E= 2L\,,\qquad \Delta_\xi C =0$$
in the scalar fields of the decomposition (\ref{HelmH}) of $H$,
$$\Delta_\xi B^\alpha = \dot L^\alpha \,,\qquad \Delta_\xi F^\alpha = - L^\alpha \,,\qquad \Delta_\xi E^\alpha = L^\alpha \,,\qquad\Delta_\xi C^\alpha = 0$$
in the solenoidal vector fields and 
$$\Delta_\xi F^{\alpha\beta} = 0\,,\qquad\Delta_\xi E_{\alpha\beta} = 0\,,\qquad\Delta_\xi C^{\alpha{}}_\beta = 0$$
in the transverse tracefree tensor fields.
Thus one finds a set of 11 gauge invariant quantities
$$
\mathscr{J}_1 = E + F \,,\qquad\quad \mathscr{J}_{2} = \widetilde E + \widetilde F \,,\qquad\quad \mathscr{J}_{3} = C\,,\qquad\quad \mathscr{J}_{4} = A +\dot B + \tfrac{1}{2} \ddot F \,,\qquad\quad \mathscr{J}_{5}=\widetilde E - \widetilde F + \tfrac{2}{3} \Delta F\,,
$$
$$\mathscr{J}_6{}^\alpha := F^\alpha + E^\alpha\,,\qquad\quad \mathscr{J}_7{}^\alpha:= B^\alpha - \dot E^\alpha\,,\qquad \quad\mathscr{J}_8{}^\alpha := C^\alpha\,,$$
$$\mathscr{J}_9{}^{\alpha\beta} := F^{\alpha\beta}\,, \qquad\quad \mathscr{J}_{10}{}^{\alpha\beta} := E^{\alpha\beta}\,, \qquad\quad \mathscr{J}_{11}{}^{\alpha\beta} := C^{\alpha\beta}\,,$$
for scalar, vector and tensor perturbations, respectively.
These are simplified by fixing a gauge, such as is obtained by letting  $L:=\tfrac{1}{2} F$, $T := B + \tfrac{1}{2} \dot F$ and $L^\alpha := - E^\alpha$, with the effect of fixing
$$E^\alpha \stackrel{*}{=} 0\,,\qquad F\stackrel{*}{=}0\,,\qquad B\stackrel{*}{=}0\,.$$
In this gauge, whose application is indicated by the asterisk $*$, we find the following decomposition of the configuration variables as well as the lapse and shift 
\begin{eqnarray*}
\varphi^{\Aa} &\stackrel{*}{=}& \I^{\Aa}{}_{\alpha\beta} \Big[ \, \widetilde{F} \,\gamma^{\alpha\beta}  + 2\partial^{(\alpha} F^{\beta)} +  F^{\alpha\beta}  \Big] \,,\\
\varphi^{\Ab} &\stackrel{*}{=}& \I^{\Ab}{}^{\alpha\beta} \Big[\, \widetilde{E}\, \gamma_{\alpha\beta} + \Delta_{\alpha\beta}E +E_{\alpha\beta}    \Big]\,,\\
\varphi^{\Ac} &\stackrel{*}{=}& \I^{\Ac}{}_{\alpha}{}^{\beta} \, \Big[ \Delta^\alpha{}_\beta C + \partial^{\alpha}C_{\beta} +  \partial_{\beta}C^{\alpha} +  C^\alpha{}_\beta \, \Big]+ \mathcal{O}(2)\,,\\
N^\alpha &\stackrel{*}{=}& B^\alpha\,,\\
N &\stackrel{*}{=}& 1 + A\,
\end{eqnarray*}
into scalar, solenoidal vector and transverse tracefree tensors, which we will put to use when writing down the linearized gravitational equations of motion in the following subsection.


\newcommand\DefineEoMCoeff[2]{%
	\expandafter\newcommand\csname #1\endcsname{#2}%
}
\newcommand\EoMCoeff[1]{%
	\csname #1\endcsname
}

\DefineEoMCoeff{ss1}{+\kappa_1}
\DefineEoMCoeff{ss1wo}{\kappa_1}
\DefineEoMCoeff{ss2}{+\kappa_2}
\DefineEoMCoeff{ss3}{+\kappa_3}
\DefineEoMCoeff{ss4}{-\kappa_3}
\DefineEoMCoeff{ss5}{+\kappa_4}
\DefineEoMCoeff{ss6}{+\kappa_5}
\DefineEoMCoeff{ss7}{+\kappa_6}
\DefineEoMCoeff{ss8}{+\kappa_7}
\DefineEoMCoeff{ss9}{+\kappa_8}
\DefineEoMCoeff{ss10}{+\left(2 \kappa_7-6 \kappa_1-6 \kappa_2-2 \kappa_8\right)} 
\DefineEoMCoeff{ss11}{+\left(\kappa_1-\kappa_7+\kappa_8\right)}
\DefineEoMCoeff{ss11wo}{\left(\kappa_1-\kappa_7+\kappa_8\right)}
\DefineEoMCoeff{ss12}{+\left(\frac{\kappa_7}{3}-\kappa_1-\frac{\kappa_8}{3}\right)}
\DefineEoMCoeff{ss13}{+\kappa_3}
\DefineEoMCoeff{ss14}{ -\kappa_3}
\DefineEoMCoeff{ss15}{+\kappa_4}
\DefineEoMCoeff{ss16}{+\kappa_5}
\DefineEoMCoeff{ss17}{+\kappa_6}
\DefineEoMCoeff{ss18}{+\left(3 \kappa_7-6 \kappa_1-6 \kappa_2-4 \kappa_8\right)} 
\DefineEoMCoeff{ss19}{+\left(2 \kappa_7-3 \kappa_1-3 \kappa_2-3 \kappa_8\right)} 
\DefineEoMCoeff{ss20}{+\left(2 \kappa_7-2 \kappa_8\right)} 
\DefineEoMCoeff{ss21}{+\kappa_3}
\DefineEoMCoeff{ss21wo}{\kappa_3}
\DefineEoMCoeff{ss22}{ -\kappa_3}
\DefineEoMCoeff{ss23}{+\kappa_9}
\DefineEoMCoeff{ss24}{ -\kappa_9}
\DefineEoMCoeff{ss25}{ -\kappa_4}
\DefineEoMCoeff{ss26}{+\kappa_6}
\DefineEoMCoeff{ss27}{-4 \kappa_5} 
\DefineEoMCoeff{ss28}{+\kappa_{10}}
\DefineEoMCoeff{ss28wo}{\kappa_{10}}
\DefineEoMCoeff{ss29}{+\left(16 \kappa_1+16 \kappa_2-\frac{8 \kappa_7}{3}+\frac{16 \kappa_8}{3}-\kappa_{10}\right)} 
\DefineEoMCoeff{ss30}{+\left(2 \kappa_7-12 \kappa_1-12 \kappa_2-4 \kappa_8+\kappa_{10}\right)} 
\DefineEoMCoeff{ss31}{+\left(20 \kappa_1+20 \kappa_2-4 \kappa_7+8 \kappa_8-\kappa_{10}\right)}
\DefineEoMCoeff{ss32}{+\kappa_{11}}
\DefineEoMCoeff{ss33}{+\kappa_{11}}
\DefineEoMCoeff{ss34}{+\left(8 \kappa_1+8 \kappa_2-\frac{4 \kappa_7}{3}+\frac{8 \kappa_8}{3}\right)} 
\DefineEoMCoeff{ss35}{+\left(-\frac{4 \kappa_1}{3}-\frac{4 \kappa_2}{3}+\frac{2 \kappa_7}{3}-\frac{8 \kappa_8}{9}\right)}
\DefineEoMCoeff{ss36}{+\left(2 \kappa_7-12 \kappa_1-12 \kappa_2-4 \kappa_8+\kappa_{10}\right)} 
\DefineEoMCoeff{ss36wo}{\left(2 \kappa_7-12 \kappa_1-12 \kappa_2-4 \kappa_8+\kappa_{10}\right)} 
\DefineEoMCoeff{ss37}{+\left(20 \kappa_1+20 \kappa_2-4 \kappa_7+8 \kappa_8-\kappa_{10}\right)} 
\DefineEoMCoeff{ss38}{+\left(4 \kappa_7-18 \kappa_1-18 \kappa_2-8 \kappa_8+\kappa_{10}\right)} 
\DefineEoMCoeff{ss39}{+\left(22 \kappa_1+22 \kappa_2-\frac{16 \kappa_7}{3}+\frac{32 \kappa_8}{3}-\kappa_{10}\right)}
\DefineEoMCoeff{ss40}{+\kappa_{11}}
\DefineEoMCoeff{ss41}{+\kappa_{11}}
\DefineEoMCoeff{ss42}{+\left(4 \kappa_1+4 \kappa_2-\frac{4 \kappa_7}{3}+\frac{8 \kappa_8}{3}\right)} 
\DefineEoMCoeff{ss43}{+\left(-\frac{2 \kappa_1}{3}-\frac{2 \kappa_2}{3}+\frac{4 \kappa_7}{9}-\frac{2 \kappa_8}{3}\right)}
\DefineEoMCoeff{ss44}{+\left(24 \kappa_1+24 \kappa_2-4 \kappa_7+8 \kappa_8\right)} 
\DefineEoMCoeff{ss44wo}{\left(24 \kappa_1+24 \kappa_2-4 \kappa_7+8 \kappa_8\right)} 
\DefineEoMCoeff{ss45}{+\left(12 \kappa_1+12 \kappa_2-4 \kappa_7+8 \kappa_8\right)}
\DefineEoMCoeff{ss46}{+\left(\frac{4 \kappa_7}{3}-\frac{4 \kappa_8}{3}\right)}

\DefineEoMCoeff{tt1}{+\left(-2 \kappa_1-3 \kappa_2+\kappa_7-\kappa_8\right)}
\DefineEoMCoeff{tt1wo}{\left(-2 \kappa_1-3 \kappa_2+\kappa_7-\kappa_8\right)}
\DefineEoMCoeff{tt2}{+\left(-3 \kappa_2+\kappa_7-\kappa_8+\kappa_9\right)} 
\DefineEoMCoeff{tt3}{+\kappa_1}
\DefineEoMCoeff{tt4}{+\left(\kappa_9-3 \kappa_2\right)}
\DefineEoMCoeff{tt5}{+\kappa_3}
\DefineEoMCoeff{tt6}{-\kappa_3}
\DefineEoMCoeff{tt7}{+\left(2 \kappa_1+6 \kappa_2-2 \kappa_7+2 \kappa_8-\kappa_9\right)} 
\DefineEoMCoeff{tt8}{+\kappa_4}
\DefineEoMCoeff{tt9}{+2 \kappa_4} 
\DefineEoMCoeff{tt10}{+2 \kappa_4} 
\DefineEoMCoeff{tt11}{+\kappa_5}
\DefineEoMCoeff{tt12}{+\kappa_5}
\DefineEoMCoeff{tt13}{+\kappa_6}
\DefineEoMCoeff{tt14}{+\kappa_1}
\DefineEoMCoeff{tt14wo}{\kappa_1}
\DefineEoMCoeff{tt15}{+\left(\kappa_9-3 \kappa_2\right)}
\DefineEoMCoeff{tt16}{+\left(\kappa_1-\kappa_7+\kappa_8\right)}
\DefineEoMCoeff{tt17}{+\left(3 \kappa_1-\kappa_7+\kappa_8+\kappa_9\right)} 
\DefineEoMCoeff{tt18}{+\kappa_3}
\DefineEoMCoeff{tt19}{-\kappa_3}
\DefineEoMCoeff{tt20}{+\left(-4 \kappa_1+2 \kappa_7-2 \kappa_8-\kappa_9\right)} 
\DefineEoMCoeff{tt21}{+\kappa_4}
\DefineEoMCoeff{tt22}{+2 \kappa_4} 
\DefineEoMCoeff{tt23}{+2 \kappa_4} 
\DefineEoMCoeff{tt24}{+\kappa_5}
\DefineEoMCoeff{tt25}{+\kappa_5}
\DefineEoMCoeff{tt26}{+\kappa_6}
\DefineEoMCoeff{tt27}{+\kappa_3} 
\DefineEoMCoeff{tt27wo}{\kappa_3} 
\DefineEoMCoeff{tt28}{-\kappa_3}
\DefineEoMCoeff{tt29}{+\kappa_3}
\DefineEoMCoeff{tt30}{-\kappa_3}
\DefineEoMCoeff{tt31}{+\kappa_9}
\DefineEoMCoeff{tt32}{+\left(4 \kappa_1-12 \kappa_2+3 \kappa_9\right)} 
\DefineEoMCoeff{tt33}{+\left(-2 \kappa_1-6 \kappa_2+2 \kappa_7-2 \kappa_8+\kappa_9\right)} 
\DefineEoMCoeff{tt34}{+\left(4 \kappa_1-2 \kappa_7+2 \kappa_8+\kappa_9\right)} 
\DefineEoMCoeff{tt35}{-\kappa_4}
\DefineEoMCoeff{tt36}{-\kappa_4}
\DefineEoMCoeff{tt37}{+8 \kappa_4} 
\DefineEoMCoeff{tt38}{+\kappa_6}
\DefineEoMCoeff{tt39}{+\kappa_6}
\DefineEoMCoeff{tt40}{-4 \kappa_5} 

\DefineEoMCoeff{vv1}{+\left(-4 \kappa_1-6 \kappa_2+2 \kappa_7-2 \kappa_8\right)} 
\DefineEoMCoeff{vv1wo}{\left(-4 \kappa_1-6 \kappa_2+2 \kappa_7-2 \kappa_8\right)} 
\DefineEoMCoeff{vv2}{+ \frac{\kappa_9}{2}}
\DefineEoMCoeff{vv3}{+ 2 \kappa_3} 
\DefineEoMCoeff{vv4}{-2 \kappa_3} 
\DefineEoMCoeff{vv5}{+\left(2 \kappa_1+6 \kappa_2-2 \kappa_7+2 \kappa_8-\kappa_9\right)} 
\DefineEoMCoeff{vv6}{+2 \kappa_4} 
\DefineEoMCoeff{vv7}{+2 \kappa_4} 
\DefineEoMCoeff{vv8}{+2 \kappa_5} 
\DefineEoMCoeff{vv9}{+2 \kappa_6} 
\DefineEoMCoeff{vv10}{+\left(-6 \kappa_1-6 \kappa_2+2 \kappa_7-2 \kappa_8\right)} 
\DefineEoMCoeff{vv11}{+2 \kappa_1} 
\DefineEoMCoeff{vv11wo}{2 \kappa_1} 
\DefineEoMCoeff{vv12}{+\frac{\kappa_9}{2}}
\DefineEoMCoeff{vv13}{+2 \kappa_3} 
\DefineEoMCoeff{vv14}{-2 \kappa_3} 
\DefineEoMCoeff{vv15}{+\left(-4 \kappa_1+2 \kappa_7-2 \kappa_8-\kappa_9\right)} 
\DefineEoMCoeff{vv16}{+2 \kappa_4} 
\DefineEoMCoeff{vv17}{+2 \kappa_4} 
\DefineEoMCoeff{vv18}{+2 \kappa_5} 
\DefineEoMCoeff{vv19}{+2 \kappa_6} 
\DefineEoMCoeff{vv20}{+\left(2 \kappa_7-2 \kappa_8\right)} 
\DefineEoMCoeff{vv21}{+2 \kappa_3} 
\DefineEoMCoeff{vv21wo}{2 \kappa_3} 
\DefineEoMCoeff{vv22}{-2 \kappa_3} 
\DefineEoMCoeff{vv23}{+2 \kappa_9} 
\DefineEoMCoeff{vv24}{+\left(2 \kappa_1-6 \kappa_2\right)} 
\DefineEoMCoeff{vv25}{+\left(-2 \kappa_1-6 \kappa_2+2 \kappa_7-2 \kappa_8+\kappa_9\right)} 
\DefineEoMCoeff{vv26}{-2 \kappa_4} 
\DefineEoMCoeff{vv27}{+8 \kappa_4} 
\DefineEoMCoeff{vv28}{+2 \kappa_6} 
\DefineEoMCoeff{vv29}{-8 \kappa_5} 
\DefineEoMCoeff{vv30}{+\left(-6 \kappa_1-6 \kappa_2+4 \kappa_7-4 \kappa_8\right)} 
\DefineEoMCoeff{vv31}{+\left(-6 \kappa_1-6 \kappa_2+2 \kappa_7-2 \kappa_8\right)} 
\DefineEoMCoeff{vv31wo}{\left(-6 \kappa_1-6 \kappa_2+2 \kappa_7-2 \kappa_8\right)} 
\DefineEoMCoeff{vv32}{+\left(-6 \kappa_1-6 \kappa_2\right)} 
\DefineEoMCoeff{vv33}{+\left(6 \kappa_1+6 \kappa_2-4 \kappa_7+4 \kappa_8\right)} 


\subsection{Gauge-fixed linearized equations of motion}
We now display the final result of this article, namely the linearized gravitational equations of motion for scalar, solenoidal vector and transverse tracefree tensor fields sourced by the birefringent electrodynamics (\ref{GLED}), whose gravitational closure gave rise to the dynamics of the geometry in the first place. The equations of motion in section IV.E of \cite{DSSW} neatly decompose into irreducible blocks and are displayed in the $E^\alpha=F=B=0$ gauge devised in the previous subsection. Moreover, since it turns out that the gravitational equations of motion only contain eleven linearly independent combinations $\kappa_1, \dots, \kappa_{11}$ of the constants $g_1, g_2, \dots, g_{22}$ that were still present in the Lagrangian, we directly employ the former and refer to table \ref{tabkappa} for their definition in terms of the latter. 

One may, of course, add any further matter, as long as it gives rise to the same principal tensor as birefringent electrodynamics. Only if one wishes to add matter with a different principal tensor, one must start the entire gravitational closure procedure again, utilizing the then different overall principal tensor for all the matter one stipulates. So in order to use the gravitational field equations derived in this article, we will assume that we are given matter with Hamiltonian $H_{\textrm{\tiny matter}}[\Psi; \varphi,N,\vec{N}]$, where $\Psi$ denotes any matter fields or even point particle trajectories, such that the principal tensor or point particle dispersion relation be given by (\ref{Ppoly}). In any case, we employ the following unique decompositions of the variations of the matter Hamiltonian with respect to the geometric configuration variables as well as lapse and shift,
\begin{eqnarray*}
	 \mathcal{I}^{\overline{A}}{}_{\alpha\beta} \frac{\delta H_{\textrm{\tiny matter}}}{\delta\varphi^{\overline{A}}} &=:& \left[\frac{\delta H_\textrm{\tiny matter}}{\delta \overline\varphi}\right]^{(TT)}_{\alpha\beta}   
	\!\!+ 2 \partial_{(\alpha} \left[\frac{\delta H_\textrm{\tiny matter}}{\delta \overline\varphi}\right]^{(V)}_{\beta)} 
	\!\!+ \gamma_{\alpha\beta} \left[\frac{\delta H_\textrm{\tiny matter}}{\delta \overline\varphi}\right]^{(S_\textrm{tr})}
	\!\!+ \Delta_{\alpha\beta} \left[\frac{\delta H_\textrm{\tiny matter}}{\delta \overline\varphi}\right]^{(S_\textrm{tr-free})}\!,\\
	\mathcal{I}^{\overline{\overline{A}}}{}_{\alpha\beta} \frac{\delta H_{\textrm{\tiny matter}}}{\delta\varphi^{\overline{\overline{A}}}} &=:&  \left[\frac{\delta H_\textrm{\tiny matter}}{\delta \overline{\overline\varphi}}\right]^{(TT)}_{\alpha\beta}   
	\!\!+ 2 \partial_{(\alpha} \left[\frac{\delta H_\textrm{\tiny matter}}{\delta \overline{\overline\varphi}}\right]^{(V)}_{\beta)} 
	\!\!+ \gamma_{\alpha\beta} \left[\frac{\delta H_\textrm{\tiny matter}}{\delta \overline{\overline\varphi}}\right]^{(S_\textrm{tr})}
	\!\!+ \Delta_{\alpha\beta} \left[\frac{\delta H_\textrm{\tiny matter}}{\delta \overline{\overline\varphi}}\right]^{(S_\textrm{tr-free})}\!,\\ 
\mathcal{I}^{\overline{\overline{\overline{A}}}}{}_{\alpha\beta} \frac{\delta H_{\textrm{\tiny matter}}}{\delta\varphi^{\overline{\overline{\overline{A}}}}} &=:& \left[\frac{\delta H_\textrm{\tiny matter}}{\delta \overline{\overline{\overline\varphi}}}\right]^{(TT)}_{\alpha\beta}   
	\!\!+ 2 \partial_{(\alpha} \left[\frac{\delta H_\textrm{\tiny matter}}{\delta \overline{\overline{\overline\varphi}}}\right]^{(V)}_{\beta)} 
	\!\!+ \Delta_{\alpha\beta} \left[\frac{\delta H_\textrm{\tiny matter}}{\delta\overline{\overline{\overline\varphi}}}\right]^{(S_\textrm{tr-free})} \!,\\
	 \frac{\delta H_{\textrm{\tiny matter}}}{\delta N^\alpha} &=:& \left[\frac{\delta H_{\textrm{\tiny matter}}}{\delta \vec{N}}\right]^{(V)}_{\alpha}
	+ \partial_\alpha  \left[\frac{\delta H_\textrm{\tiny matter}}{\delta\vec{N}}\right]^{(S)} \!,\\
	\frac{\delta H_{\textrm{\tiny matter}}}{\delta N} &=:& \left[\frac{\delta H_{\textrm{\tiny matter}}}{\delta N}\right]^{(S)}\!,
\end{eqnarray*}
where indices on the left and the right hand side have been raised and lowered with the Euclidean background metric $\gamma$. The linearization of the various components on the right hand side will provide the left hand side of the gravitational field equations, where the matter fields are taken to be of comparable size to the geometric perturbations $\varphi^A$, $A$ and $B^\alpha$. 

First, we display the equations of motion for the scalar modes $\widetilde E, \widetilde F, E, C$ and $A$, which come as the five evolution equations 
\begin{eqnarray*}
	\left[\frac{\delta H_\textrm{matter}}{\delta \overline\varphi}\right]^{(S_\textrm{tr-free})} \!\!\! &\stackrel{*}{=}& 
		\EoMCoeff{ss1wo} \ddot{E}	
		\EoMCoeff{ss2} \Delta E
	    \EoMCoeff{ss3} \ddot{C} 
	    \EoMCoeff{ss4} \Delta C
	    \EoMCoeff{ss5} \dot{C} 
	    \EoMCoeff{ss6} E 
	    \EoMCoeff{ss7} C
	    \EoMCoeff{ss8} \widetilde{F} 
	    \EoMCoeff{ss9} \widetilde{E}\\
	    & & \EoMCoeff{ss10} A\,,
	    \\
	\left[\frac{\delta H_\textrm{matter}}{\delta \overline\varphi}\right]^{(S_\textrm{tr})} \,\,\,\,\, &\stackrel{*}{=}& 
		\EoMCoeff{ss28wo} \ddot{\widetilde{F}} 
		\EoMCoeff{ss29} \Delta\widetilde{F}				
		\EoMCoeff{ss32} \widetilde{F} 
		\\ & & 
		\EoMCoeff{ss30} \ddot{\widetilde{E}} 
		\EoMCoeff{ss31} \Delta\widetilde{E}
		\\ & &\EoMCoeff{ss33} \widetilde{E} 
		\EoMCoeff{ss34} \Delta A 
		\EoMCoeff{ss35} \Delta\Delta E\,,
		\\
	\left[\frac{\delta H_\textrm{matter}}{\delta \overline{\overline\varphi}}\right]^{(S_\textrm{tr-free})} \!\!\! &\stackrel{*}{=}& 
		\EoMCoeff{ss11wo} \ddot{E} 
		\EoMCoeff{ss12} \Delta E 
		\EoMCoeff{ss16} E 
		\\ & &	
		\EoMCoeff{ss13} \ddot{C} 
		\EoMCoeff{ss14} \Delta C
		\EoMCoeff{ss15} \dot{C} 
		\EoMCoeff{ss17} C
		\\ & &
		\EoMCoeff{ss18} \widetilde{F} 
		\EoMCoeff{ss19} \widetilde{E} 
		\EoMCoeff{ss20} A\,,
		\\
	\left[\frac{\delta H_\textrm{matter}}{\delta \overline{\overline\varphi}}\right]^{(S_\textrm{tr})} \,\,\,\,\, &\stackrel{*}{=}& 
		\EoMCoeff{ss36wo} \ddot{\widetilde{F}} 
		\EoMCoeff{ss37} \Delta \widetilde{F}
		\\ & & 
		\EoMCoeff{ss38} \ddot{\widetilde{E}} 
        \EoMCoeff{ss43} \Delta\Delta E
        \\ & &
		\EoMCoeff{ss39} \Delta \widetilde{E} 
        \EoMCoeff{ss42} \Delta A 
        \\ & &
        \EoMCoeff{ss40} \widetilde{F} 
        \EoMCoeff{ss41} \widetilde{E} \,,
		\\
	\left[\frac{\delta H_\textrm{matter}}{\delta \overline{\overline{\overline\varphi}}}\right]^{(S_\textrm{tr-free})}  \!\!\! &\stackrel{*}{=}& 
		\EoMCoeff{ss21wo} \ddot{E} 
		\EoMCoeff{ss22} \Delta E 
		\EoMCoeff{ss25} \dot{E} 
		\EoMCoeff{ss26} E 
		\EoMCoeff{ss23} \ddot{C} 
		\EoMCoeff{ss24} \Delta C 
		\EoMCoeff{ss27} C\,,
\end{eqnarray*}
where the three generic evolution equations were further separated into trace and tracefree parts, plus the two constraint equations
\begin{eqnarray*}
	\left[\frac{\delta H_\textrm{matter}}{\delta \vec{N}}\right]^{(S)} \,\,\,\,\,\,\,\, &\stackrel{*}{=}& 
		\EoMCoeff{ss44wo} \dot{\widetilde{F}\,} 
		\EoMCoeff{ss45} \dot{\widetilde{E}\,}  \\ & &
		\EoMCoeff{ss46} \Delta \dot E\,,
		\\
	\left[\frac{\delta H_\textrm{matter}}{\delta N}\right]^{(S)} \,\,\,\,\,\,\,\, &\stackrel{*}{=}& 
		\EoMCoeff{ss44wo} \Delta\widetilde{F} 
		\EoMCoeff{ss45} \Delta\widetilde{E} \qquad\qquad\\ & &
		\EoMCoeff{ss46} \Delta\Delta E\,
\end{eqnarray*}
that restrict the initial data that can be evolved by the five evolution equations above. The asterisk simply denotes usage of our particular gauge.  
\newpage 
Similarly, we have the three evolution equations plus one constraint for the solenoidal vector fields $F^\alpha, C^\alpha$ and $B^\alpha$
\begin{eqnarray*}
	2 \partial_{(\alpha} \left[\frac{\delta H_\textrm{matter}}{\delta \overline\varphi}\right]^{(V)}_{\beta)}  &\stackrel{*}{=}& 
		\EoMCoeff{vv1wo} \partial_{(\alpha} \ddot{F}_{\beta)}  
		\EoMCoeff{vv2} \Delta \partial_{(\alpha} F_{\beta)}
		\EoMCoeff{vv7} \epsilon{}_{(\alpha}{}^{\gamma\mu} \partial_{\beta)} F_{\gamma,\mu} 
		\EoMCoeff{vv8} \partial_{(\alpha} F_{\beta)} 
		\\ & &
		\EoMCoeff{vv3} \partial_{(\alpha} \ddot{C}_{\beta)} 
		\EoMCoeff{vv4} \Delta \partial_{(\alpha} C_{\beta)} 
		\EoMCoeff{vv5} \epsilon{}_{(\alpha}{}^{\gamma\mu} \partial_{\beta)} \dot{C}_{\gamma,\mu}
        \\ & &
		\EoMCoeff{vv6} \partial_{(\alpha} \dot{C}_{\beta)}
        \EoMCoeff{vv9} \partial_{(\alpha} C_{\beta)}
		\EoMCoeff{vv10} \dot{B}_{(\alpha,\beta)} \,,
		\\
	2 \partial_{(\alpha} \left[\frac{\delta H_\textrm{matter}}{\delta \overline{\overline\varphi}}\right]^{(V)}_{\beta)}  &\stackrel{*}{=}& 
		\EoMCoeff{vv11wo} \partial_{(\alpha} \ddot{F}_{\beta)} 
		\EoMCoeff{vv12} \Delta \partial_{(\alpha} F_{\beta)} 
		\EoMCoeff{vv17} \epsilon{}_{(\alpha}{}^{\gamma\mu} \partial_{\beta)} F_{\gamma,\mu}
		\EoMCoeff{vv18} \partial_{(\alpha} F_{\beta)} 
		\\ & &
		\EoMCoeff{vv13} \partial_{(\alpha} \ddot{C}_{\beta)} 
		\EoMCoeff{vv14} \Delta \partial_{(\alpha} C_{\beta)} 
		\EoMCoeff{vv15} \epsilon{}_{(\alpha}{}^{\gamma\mu} \partial_{\beta)} \dot{C}_{\gamma,\mu} 
        \\ & &
		\EoMCoeff{vv16} \partial_{(\alpha} \dot{C}_{\beta)}
		\EoMCoeff{vv19} \partial_{(\alpha} C_{\beta)}
		\EoMCoeff{vv20} \partial_{(\alpha} \dot{B}_{\beta)}\,, 
		\\
	2 \partial_{(\alpha} \left[\frac{\delta H_\textrm{matter}}{\delta \overline{\overline{\overline\varphi}}}\right]^{(V)}_{\beta)}  &\stackrel{*}{=}& 
		\EoMCoeff{vv21wo} \partial_{(\alpha} \ddot{F}_{\beta)}
		\EoMCoeff{vv22} \Delta \partial_{(\alpha} F_{\beta)}
		\EoMCoeff{vv25} \epsilon{}_{(\alpha}{}^{\gamma\mu} \partial_{\beta)} \dot{F}_{\gamma,\mu}
        \\ & &
		\EoMCoeff{vv26} \partial_{(\alpha} \dot{F}_{\beta)}
		\EoMCoeff{vv28} \partial_{(\alpha} F_{\beta)}
		\EoMCoeff{vv23} \partial_{(\alpha} \ddot{C}_{\beta)}
		\EoMCoeff{vv24} \Delta \partial_{(\alpha} C_{\beta)} 
        \\ & &
		\EoMCoeff{vv27} \epsilon{}_{(\alpha}{}^{\gamma\mu} \partial_{\beta)} C_{\gamma,\mu} 
		\EoMCoeff{vv29} \partial_{(\alpha} C_{\beta)}
		\EoMCoeff{vv30} \epsilon{}_{(\alpha}{}^{\gamma\mu} \partial_{\beta)} B_{\gamma,\mu} \,,
		\\
	\left[\frac{\delta H_\textrm{matter}}{\delta \vec{N}}\right]^{(V)}_{\alpha}  &\stackrel{*}{=}& \EoMCoeff{vv31wo} \Delta \dot{F}_{\alpha}
		\EoMCoeff{vv32} \Delta B_{\alpha} 
        \\ & &
		\EoMCoeff{vv33} \epsilon{}_{\alpha}{}^{\gamma\mu} \Delta C_{\gamma,\mu}\,.
\end{eqnarray*}

Finally, we display the tensor equations. The transverse traceless tensor perturbations $F^{\alpha\beta}$, $E^{\alpha\beta}$ and $C^{\alpha\beta}$ are all gauge invariant quantities and are determined by the three evolution equations
\begin{eqnarray*}
	\left[\frac{\delta H_\textrm{matter}}{\delta \overline\varphi}\right]^{(TT)}_{\alpha\beta}   &\stackrel{*}{=}&
		\EoMCoeff{tt1wo} \ddot{F}_{\alpha\beta}
		\EoMCoeff{tt2} \Delta F_{\alpha\beta}
		\\& &
		\EoMCoeff{tt9} \epsilon{}_{(\alpha}{}^{\gamma\mu} F_{\beta)\gamma,\mu}
		\EoMCoeff{tt11} F_{\alpha\beta}
		\EoMCoeff{tt3} \ddot{E}_{\alpha\beta}
		\EoMCoeff{tt4} \Delta E_{\alpha\beta}
		\\ & &
		\EoMCoeff{tt10} \epsilon{}_{(\alpha}{}^{\gamma\mu} E_{\beta)\gamma,\mu}
		\EoMCoeff{tt12} E_{\alpha\beta}
		\EoMCoeff{tt7} \epsilon{}_{(\alpha}{}^{\gamma\mu} \dot{C}_{\beta)\gamma,\mu}
		\\ & &
		\EoMCoeff{tt5} \ddot{C}_{\alpha\beta}
		\EoMCoeff{tt6} \Delta C_{\alpha\beta}
		\EoMCoeff{tt8} \dot{C}_{\alpha\beta}		
		\EoMCoeff{tt13} C_{\alpha\beta}\,, 
		\\
	\left[\frac{\delta H_\textrm{matter}}{\delta \overline{\overline\varphi}}\right]^{(TT)}_{\alpha\beta}  &\stackrel{*}{=}&
		\EoMCoeff{tt14wo} \ddot{F}_{\alpha\beta}
		\EoMCoeff{tt15} \Delta F_{\alpha\beta}
		\EoMCoeff{tt22} \epsilon{}_{(\alpha}{}^{\gamma\mu} F_{\beta)\gamma,\mu}
		\EoMCoeff{tt24} F_{\alpha\beta} 
		\EoMCoeff{tt16} \ddot{E}_{\alpha\beta}
		\\ & &
		\EoMCoeff{tt17} \Delta E_{\alpha\beta}
		\EoMCoeff{tt23} \epsilon{}_{(\alpha}{}^{\gamma\mu} E_{\beta)\gamma,\mu}
		\EoMCoeff{tt25} E_{\alpha\beta}
		\EoMCoeff{tt18} \ddot{C}_{\alpha\beta}
		\\ & &
		\EoMCoeff{tt19} \Delta C_{\alpha\beta}
		\EoMCoeff{tt20} \epsilon{}_{(\alpha}{}^{\gamma\mu} \dot{C}_{\beta)\gamma,\mu}
		\EoMCoeff{tt21} \dot{C}_{\alpha\beta}
		\EoMCoeff{tt26} C_{\alpha\beta} \,,
		\\
	\left[\frac{\delta H_\textrm{matter}}{\delta \overline{\overline{\overline\varphi}}}\right]^{(TT)}_{\alpha\beta}   &\stackrel{*}{=}&
		\EoMCoeff{tt27wo} \ddot{F}_{\alpha\beta}
		\EoMCoeff{tt28} \Delta F_{\alpha\beta}
		\EoMCoeff{tt33} \epsilon{}_{(\alpha}{}^{\gamma\mu} \dot{F}_{\beta)\gamma,\mu}
		\\ & &
		\EoMCoeff{tt35} \dot{F}_{\alpha\beta}
		\EoMCoeff{tt38} F_{\alpha\beta}
               \EoMCoeff{tt29} \ddot{E}_{\alpha\beta}
		\EoMCoeff{tt30} \Delta E_{\alpha\beta}
		\\ & &
		\EoMCoeff{tt34} \epsilon{}_{(\alpha}{}^{\gamma\mu} \dot{E}_{\beta)\gamma,\mu}
		\EoMCoeff{tt36} \dot{E}_{\alpha\beta} 
		\EoMCoeff{tt39} E_{\alpha\beta}
		\\ & &
		\EoMCoeff{tt31} \ddot{C}_{\alpha\beta}
		\EoMCoeff{tt32} \Delta C_{\alpha\beta}
		\EoMCoeff{tt37} \epsilon{}_{(\alpha}{}^{\gamma\mu} C_{\beta)\gamma,\mu}
		\EoMCoeff{tt40} C_{\alpha\beta}\,.
\end{eqnarray*}

\setlength{\tabcolsep}{12pt}
\begin{table}
\noindent\begin{tabularx}{\textwidth}{cl}
 \hline
	$\textsc{Coefficient in EOM}$ & \scshape Combination of coefficients in Lagrangian \\
	\hline
	$\kappa_1$ & $-\frac{8}{3}g_2+\frac{2}{3}g_4+\frac{7}{6}g_6+g_9-2 g_{10}+g_{11}-\frac{1}{3}g_7-\frac{13}{6}g_3$ \\
	$\kappa_2$ & $\frac{8}{3}g_2+\frac{5}{3}g_3+\frac{8}{3}g_7+2 g_{10}-g_{11}-\frac{2}{3}g_4 -\frac{4}{3}g_6-\frac{2}{3}g_8-\frac{1}{3}g_9$ \\
	$\kappa_3$ & $-2 g_{12}$ \\
	$\kappa_4$ & $g_{17}$ \\
	$\kappa_5$ & $g_{21}$ \\
	$\kappa_6$ & $g_{22}$ \\
	$\kappa_7$ & $-g_2+g_3+\frac{1}{4}g_4+\frac{3}{2}g_6-8 g_7-\frac{5}{2}g_9 $ \\
	$\kappa_8$ & $-g_2+2 g_3+\frac{1}{4}g_4+\frac{3}{2}g_6-12 g_7+2 g_8-\frac{7}{2}g_9$ \\
	$\kappa_9$ & $\frac{32}{3}g_2+\frac{20}{3}g_3+\frac{16}{3}g_7-2 g_9+8 g_{10}-4 g_{11}-\frac{8}{3}g_4-\frac{14}{3}g_6$ \\
	$\kappa_{10}$ & $-6 g_1+\frac{13}{6}g_4+3 g_5+\frac{5}{3}g_6-2 g_{10}+g_{11}-\frac{26}{3}g_2-\frac{5}{3}g_3-\frac{10}{3}g_7 $ \\
	$\kappa_{11}$ & $3 g_{18}+g_{21}$ \\
	\hline
\end{tabularx}
\caption{The definition of the 11 gravitational constants that appear in the linearized gravitational equations of motion, in terms of the 22 constants appearing in the Lagrangian.\label{tabkappa}}
\end{table}

Solution of these linear equations, for the field generating matter of interest, is now a standard mathematical physics problem and will be discussed for a number of physically interesting situations in a separate article. The point we wished to explicitly illustrate in this paper is that the above linearized field equations for the fourth-rank tensor field $G$, which provides the geometric background for birefringent electrodynamics (\ref{GLED}), follows directly from the latter by gravitational closure. Thus this presents the first example of an explicit gravitational theory, other than general relativity, that has been obtained by gravitational closure of specific matter field dynamics.

\newpage
\section{Conclusions}\label{sec_conclusions}
The aim of this article was to derive, by way of the gravitational closure mechanism, the weak field gravitational dynamics for the refined geometry that underlies the most general linear extension of Maxwell electrodynamics. Had we  performed the complete closure program by starting, instead, from standard Maxwell theory on a Lorentzian metric background, we would have obtained the familiar Einstein field equations, up to  two undetermined constants of integration: the gravitational and the cosmological constant. But for the electrodynamics considered here, which admit birefringence of light in vacuo, the underlying spacetime geometry is refined and this is reflected in the geometry being encoded in a tensor field of rank four. The resulting linearized field equations, to which we restricted attention here, describe the scalar, solenoidal vector and transverse tracefree tensor perturbations of this  refined geometry around a flat non-birefringent background and feature eleven undetermined constants of integration. 

These are the most general gravitational field equations whose canonical evolution can proceed in lockstep with the evolution of the electrodynamic theory from which they are derived. 
This theoretical consistency by itself can of course not yet guarantee the physical validity of the obtained gravitational dynamics. But it allows to investigate the gravitational consequences that are equivalent to the assumption of a potential birefringence of light in vacuo. This opens up two important theoretical and experimental avenues.

First, the very search for such vacuum birefringence is now solidly underpinned by being able to quantitatively ask and answer the question where in the universe birefringence will occur and how strong it will be. 
This allows to experimentally confront the hypothesis of a physical spacetime structure admitting birefringence of matter in vacuo
in a systematic way. Indeed, both the hyperfine structure of hydrogen and quantum electrodynamical scattering amplitudes are characteristically affected by weakly birefringent gravitational fields \cite{GST}, and this will more likely lead to a direct detection of birefringence than the much coarser effect of macroscopic light ray splitting once one knows where precisely to expect an effect of which strength. Another observationally accessible \cite{Moreetal} effect is a modification of Etherington's standard relation between the apparent luminosity and angular diameter  of lensed objects \cite{SW}. 

Secondly, with the gravitational closure of birefringent electrodynamics now available, one no longer needs to observe electromagnetic radiation itself in order to draw conclusions about its potential vacuum birefringence. For instance, one can now ask and answer the question of how the refined spacetime structure, which enables the birefringence of light, affects the generation and propagation of gravitational waves.
Indeed, if the weak field equations obtained by our gravitational closure of birefringent electrodynamics were to yield observationally excluded consequences, say for the gravitational radiation off a binary, one could immediately dispense with the entire idea of birefringent electrodynamics and a correspondingly refined spacetime geometry, even without ever having observed a single light ray, let alone one that splits up in vacuo. Apart from their generation, birefringent gravitational waves propagating through the space between a pulsar and Earth will leave a distinct mark on the timing \cite{Kramer} of the signals we receive.

The linearized gravitational field equations obtained in this paper are systematically solved and studied in a separate publication for various situations of physical interest, such as vacuum solutions, the weak field around a rotating point mass and the far field of a binary source, and will yield the answers to the above questions and others. Making physically relevant symmetry assumptions, such as the presence of cosmological or spherical Killing symmetries, one can solve the pertinent closure equations for birefringent electodynamics exactly \cite{DFS}. Taking due care to not run into problems with symmetric criticality \cite{Palais,moderncriticality}, this complements the perturbative treatment pursued in this article. One is thus able to bypass the problem of a general solution of the closure equations, because any additional assumption we would have to make also in standard general realtivity in order to obtain physically relevant solutions of the gravitational field equations can be made already at the level of the closure equations. 

Nevertheless, it would be desirable to obtain the exact field equations without symmetry assumptions in order to understand also the non-linear, non-symmetric regime of the gravity theory underlying birefringent electrodynamics. One possible route to achieve this could be to emulate the bootstrap method \cite{Deser1970} that recovers the exact Einstein equations from their linearized version, starting from the linearized gravitational field equations obtained in this article. 

Beyond the technical results, the probably most interesting point illustrated by the present paper is that the gravitational closure mechanism indeed is a successful method to derive the gravitational consequences of a canonically quantizable matter model based on a tensorial spacetime geometry. In any such case, it allows to translate fundamental insights about matter into insights about gravity. Even if the spacetime is populated by a multitude of different matter fields, one single gravity theory for their combined background geometry is obtained by gravitational closure. 

\acknowledgments
The authors thank Marcus Werner, Shinji Mukohyama and Antonio de Felice for valuable remarks. JS and FPS thank the Yukawa Institute for Theoretical Physics for their hospitality and generosity during their respective three-month and two-month stays in spring and autumn  2016, where  significant parts of this work were undertaken and completed. NS thanks the Studienstiftung des deutschen Volkes for financial support.

\bibliography{references}{}

\begin{thebibliography}{24}
\expandafter\ifx\csname natexlab\endcsname\relax\def\natexlab#1{#1}\fi
\expandafter\ifx\csname bibnamefont\endcsname\relax
  \def\bibnamefont#1{#1}\fi
\expandafter\ifx\csname bibfnamefont\endcsname\relax
  \def\bibfnamefont#1{#1}\fi
\expandafter\ifx\csname citenamefont\endcsname\relax
  \def\citenamefont#1{#1}\fi
\expandafter\ifx\csname url\endcsname\relax
  \def\url#1{\texttt{#1}}\fi
\expandafter\ifx\csname urlprefix\endcsname\relax\def\urlprefix{URL }\fi
\providecommand{\bibinfo}[2]{#2}
\providecommand{\eprint}[2][]{\url{#2}}

\bibitem[{\citenamefont{Hojman et~al.}(1976)\citenamefont{Hojman, Kuchar, and
  Teitelboim}}]{HKT}
\bibinfo{author}{\bibfnamefont{S.~A.} \bibnamefont{Hojman}},
  \bibinfo{author}{\bibfnamefont{K.}~\bibnamefont{Kuchar}}, \bibnamefont{and}
  \bibinfo{author}{\bibfnamefont{C.}~\bibnamefont{Teitelboim}},
  \bibinfo{journal}{Annals Phys.} \textbf{\bibinfo{volume}{96}},
  \bibinfo{pages}{88} (\bibinfo{year}{1976}).

\bibitem[{\citenamefont{Raetzel et~al.}(2011)\citenamefont{Raetzel, Rivera, and
  Schuller}}]{RRS}
\bibinfo{author}{\bibfnamefont{D.}~\bibnamefont{Raetzel}},
  \bibinfo{author}{\bibfnamefont{S.}~\bibnamefont{Rivera}}, \bibnamefont{and}
  \bibinfo{author}{\bibfnamefont{F.~P.} \bibnamefont{Schuller}},
  \bibinfo{journal}{Phys. Rev.} \textbf{\bibinfo{volume}{D83}},
  \bibinfo{pages}{044047} (\bibinfo{year}{2011}), \eprint{1010.1369}.

\bibitem[{\citenamefont{Witte}(2014)}]{WittePhD}
\bibinfo{author}{\bibfnamefont{C.}~\bibnamefont{Witte}}, Ph.D. thesis,
  \bibinfo{school}{HU Berlin} (\bibinfo{year}{2014}).

\bibitem[{\citenamefont{Duell et~al.}(2016)\citenamefont{Duell, Schuller,
  Stritzelberger, and Wolz}}]{DSSW}
\bibinfo{author}{\bibfnamefont{M.}~\bibnamefont{Duell}},
  \bibinfo{author}{\bibfnamefont{F.~P.} \bibnamefont{Schuller}},
  \bibinfo{author}{\bibfnamefont{N.}~\bibnamefont{Stritzelberger}},
  \bibnamefont{and} \bibinfo{author}{\bibfnamefont{F.}~\bibnamefont{Wolz}}
  (\bibinfo{year}{2016}), \eprint{1611.08878}.

\bibitem[{\citenamefont{Hehl and Obukhov}(2003)}]{HehlBook}
\bibinfo{author}{\bibfnamefont{F.~W.} \bibnamefont{Hehl}} \bibnamefont{and}
  \bibinfo{author}{\bibfnamefont{Y.~N.} \bibnamefont{Obukhov}},
  \emph{\bibinfo{title}{{Foundations of Classical Electrodynamics}}}
  (\bibinfo{publisher}{Birkhäuser}, \bibinfo{year}{2003}).

\bibitem[{\citenamefont{Obukhov and Rubilar}(2002)}]{Rubilar}
\bibinfo{author}{\bibfnamefont{Y.~N.} \bibnamefont{Obukhov}} \bibnamefont{and}
  \bibinfo{author}{\bibfnamefont{G.~F.} \bibnamefont{Rubilar}},
  \bibinfo{journal}{Phys. Rev.} \textbf{\bibinfo{volume}{D66}},
  \bibinfo{pages}{024042} (\bibinfo{year}{2002}), \eprint{gr-qc/0204028}.

\bibitem[{\citenamefont{Schuller et~al.}(2010)\citenamefont{Schuller, Witte,
  and Wohlfarth}}]{SWW}
\bibinfo{author}{\bibfnamefont{F.~P.} \bibnamefont{Schuller}},
  \bibinfo{author}{\bibfnamefont{C.}~\bibnamefont{Witte}}, \bibnamefont{and}
  \bibinfo{author}{\bibfnamefont{M.~N.~R.} \bibnamefont{Wohlfarth}},
  \bibinfo{journal}{Annals Phys.} \textbf{\bibinfo{volume}{325}},
  \bibinfo{pages}{1853} (\bibinfo{year}{2010}), \eprint{0908.1016}.

\bibitem[{\citenamefont{Rivera and Schuller}(2011)}]{SR}
\bibinfo{author}{\bibfnamefont{S.}~\bibnamefont{Rivera}} \bibnamefont{and}
  \bibinfo{author}{\bibfnamefont{F.~P.} \bibnamefont{Schuller}},
  \bibinfo{journal}{Phys. Rev.} \textbf{\bibinfo{volume}{D83}},
  \bibinfo{pages}{064036} (\bibinfo{year}{2011}), \eprint{1101.0491}.

\bibitem[{\citenamefont{Pfeifer and Siemssen}(2016)}]{Pfeifer}
\bibinfo{author}{\bibfnamefont{C.}~\bibnamefont{Pfeifer}} \bibnamefont{and}
  \bibinfo{author}{\bibfnamefont{D.}~\bibnamefont{Siemssen}},
  \bibinfo{journal}{Phys. Rev.} \textbf{\bibinfo{volume}{D93}},
  \bibinfo{pages}{105046} (\bibinfo{year}{2016}), \eprint{1602.00946}.

\bibitem[{\citenamefont{Kostelecky}(2004)}]{Kostelecky}
\bibinfo{author}{\bibfnamefont{V.~A.} \bibnamefont{Kostelecky}},
  \bibinfo{journal}{Phys. Rev.} \textbf{\bibinfo{volume}{D69}},
  \bibinfo{pages}{105009} (\bibinfo{year}{2004}), \eprint{hep-th/0312310}.

\bibitem[{\citenamefont{Kostelecky and Mewes}(2009)}]{theentireposse}
\bibinfo{author}{\bibfnamefont{V.~A.} \bibnamefont{Kostelecky}}
  \bibnamefont{and} \bibinfo{author}{\bibfnamefont{M.}~\bibnamefont{Mewes}},
  \bibinfo{journal}{Phys. Rev.} \textbf{\bibinfo{volume}{D80}},
  \bibinfo{pages}{015020} (\bibinfo{year}{2009}), \eprint{0905.0031}.

\bibitem[{\citenamefont{Perlick}(2011)}]{Perlick}
\bibinfo{author}{\bibfnamefont{V.}~\bibnamefont{Perlick}}, \bibinfo{journal}{J.
  Math. Phys.} \textbf{\bibinfo{volume}{52}}, \bibinfo{pages}{042903}
  (\bibinfo{year}{2011}), \eprint{1011.2536}.

\bibitem[{\citenamefont{Palais}(1979)}]{Palais}
\bibinfo{author}{\bibfnamefont{R.~S.} \bibnamefont{Palais}},
  \bibinfo{journal}{Commun. Math. Phys.} \textbf{\bibinfo{volume}{69}},
  \bibinfo{pages}{19} (\bibinfo{year}{1979}).

\bibitem[{\citenamefont{Fels and Torre}(2002)}]{moderncriticality}
\bibinfo{author}{\bibfnamefont{M.~E.} \bibnamefont{Fels}} \bibnamefont{and}
  \bibinfo{author}{\bibfnamefont{C.~G.} \bibnamefont{Torre}},
  \bibinfo{journal}{Class. Quant. Grav.} \textbf{\bibinfo{volume}{19}},
  \bibinfo{pages}{641} (\bibinfo{year}{2002}), \eprint{gr-qc/0108033}.

\bibitem[{\citenamefont{Reiss}(2014)}]{Reiss}
\bibinfo{author}{\bibfnamefont{D.~A.} \bibnamefont{Reiss}}, \bibinfo{type}{{BSc
  thesis}}, \bibinfo{school}{University Erlangen-Nuremberg}
  (\bibinfo{year}{2014}).

\bibitem[{\citenamefont{Wierzba}(2015)}]{Wierzba}
\bibinfo{author}{\bibfnamefont{A.}~\bibnamefont{Wierzba}}, \bibinfo{type}{{BSc
  thesis}}, \bibinfo{school}{University Erlangen-Nuremberg}
  (\bibinfo{year}{2015}).

\bibitem[{\citenamefont{Stritzelberger}(2016)}]{Nthesis}
\bibinfo{author}{\bibfnamefont{N.}~\bibnamefont{Stritzelberger}},
  \bibinfo{type}{{MSc thesis}}, \bibinfo{school}{University Erlangen-Nuremberg}
  (\bibinfo{year}{2016}).

\bibitem[{\citenamefont{Schneider}(2017)}]{SchneiderMSc}
\bibinfo{author}{\bibfnamefont{J.}~\bibnamefont{Schneider}},
  \bibinfo{type}{{MSc thesis}}, \bibinfo{school}{University Erlangen-Nuremberg}
  (\bibinfo{year}{2017}).

\bibitem[{\citenamefont{Grosse-Holz et~al.}(2017)\citenamefont{Grosse-Holz,
  Schuller, and Tanzi}}]{GST}
\bibinfo{author}{\bibfnamefont{S.}~\bibnamefont{Grosse-Holz}},
  \bibinfo{author}{\bibfnamefont{F.~P.} \bibnamefont{Schuller}},
  \bibnamefont{and} \bibinfo{author}{\bibfnamefont{R.}~\bibnamefont{Tanzi}}
  (\bibinfo{year}{2017}), \eprint{1703.07183}.

\bibitem[{\citenamefont{More et~al.}(2016)\citenamefont{More, Niikura,
  Schneider, Schuller, and Werner}}]{Moreetal}
\bibinfo{author}{\bibfnamefont{S.}~\bibnamefont{More}},
  \bibinfo{author}{\bibfnamefont{H.}~\bibnamefont{Niikura}},
  \bibinfo{author}{\bibfnamefont{J.}~\bibnamefont{Schneider}},
  \bibinfo{author}{\bibfnamefont{F.~P.} \bibnamefont{Schuller}},
  \bibnamefont{and} \bibinfo{author}{\bibfnamefont{M.~C.} \bibnamefont{Werner}}
  (\bibinfo{year}{2016}), \eprint{1612.08784}.

\bibitem[{\citenamefont{Schuller and Werner}(2017)}]{SW}
\bibinfo{author}{\bibfnamefont{F.~P.} \bibnamefont{Schuller}} \bibnamefont{and}
  \bibinfo{author}{\bibfnamefont{M.~C.} \bibnamefont{Werner}},
  \bibinfo{journal}{Universe} \textbf{\bibinfo{volume}{3}}, \bibinfo{pages}{52}
  (\bibinfo{year}{2017}), \eprint{1707.01261}.

\bibitem[{\citenamefont{Becker et~al.}(2017)\citenamefont{Becker, Kramer, and
  Sesana}}]{Kramer}
\bibinfo{author}{\bibfnamefont{W.}~\bibnamefont{Becker}},
  \bibinfo{author}{\bibfnamefont{M.}~\bibnamefont{Kramer}}, \bibnamefont{and}
  \bibinfo{author}{\bibfnamefont{A.}~\bibnamefont{Sesana}}
  (\bibinfo{year}{2017}), \eprint{1705.11022}.

\bibitem[{\citenamefont{Duell et~al.}(2017)\citenamefont{Duell, Fischer, and
  Schuller}}]{DFS}
\bibinfo{author}{\bibfnamefont{M.}~\bibnamefont{Duell}},
  \bibinfo{author}{\bibfnamefont{N.}~\bibnamefont{Fischer}}, \bibnamefont{and}
  \bibinfo{author}{\bibfnamefont{F.~P.} \bibnamefont{Schuller}},
  \bibinfo{journal}{in preparation}  (\bibinfo{year}{2017}).

\bibitem[{\citenamefont{Deser}(1970)}]{Deser1970}
\bibinfo{author}{\bibfnamefont{S.}~\bibnamefont{Deser}}, \bibinfo{journal}{Gen.
  Rel. Grav.} \textbf{\bibinfo{volume}{1}}, \bibinfo{pages}{9}
  (\bibinfo{year}{1970}), \eprint{gr-qc/0411023}.

\end{thebibliography}

\appendix

\section*{Appendix: The eighty relations for expansion coefficients $\theta$, $\xi$, $\lambda$}
\setlength{\tabcolsep}{3pt}
\noindent\begin{longtable}{lcp{0.8\textwidth}}
\hline
	\textsc{Eqn} & \textsc{Order} & \textsc{Algebraic relation for expansion coefficients} \\
	\hline
	\endhead
	(C1) & $0$ & $0 = \lambda\kd{\gamma}{\mu} + 2\Ii{A}{\mu\s1}\gu{\s1\gamma}\lb{A} - 2\Iii{A}{\gamma\s1}\gd{\s1\mu}\lbb{A}$ \\
    & $\p{M}$ & $0 = \lb{M}\kd{\gamma}{\mu} + 2\Ii{A}{\mu\s1}\r{\i{\s1\gamma}{M}\lb{A} + \gu{\s1\gamma}\lbB{A}{M}} - 2\Iii{A}{\gamma\s1}\gd{\s1\mu}\lbBB{M}{A}$ \\
	 & $\pp{M}$ & $0 = \lbb{M}\kd{\gamma}{\mu} - 2\Iii{A}{\gamma\s1}\r{\ii{\s1\mu}{M}\lbb{A} + \gd{\s1\mu}\lbbBB{A}{M}} + 2\Ii{A}{\mu\s1}\gu{\s1\gamma}\lbBB{A}{M}$ \\
	 & $\ppp{M}$ & $0 = \lbbb{M}\kd{\gamma}{\mu} + \Iiii{A}{\mu}{\s1}\iii{\gamma}{\s1}{M}\lbbb{A} - \Iiii{A}{\s1}{\gamma}\iii{\s1}{\mu}{M}\lbbb{A} + 2\Ii{A}{\mu\s1}\gu{\s1\gamma}\lbBBB{A}{M}$ \\
    & & $\hphantom{=} - 2\Iii{A}{\gamma\s1}\gd{\mu\s1}\lbbBBB{A}{M}$ \\
     & $\P{M}[,\alpha]$ & $0 = 2\Ii{A}{\mu\s1}\gu{\s1\gamma}\LLd{\overline{A}}{M}{\alpha} - 2\Iii{A}{\gamma\s1}\gd{\mu\s1}\LLd{\overline{\overline{A}}}{M}{\alpha}$ \\
     & $\p{M}[,\alpha\beta]$ & $0 = \lbdd{M}{\alpha\beta}\kd{\gamma}{\mu} - 2\kd{(\alpha}{\mu}\lbdd{M}{\beta)\gamma} + 2\Ii{A}{\mu\s1}\gu{\s1\gamma}\lbBdd{A}{M}{\alpha\beta} - 2\Iii{A}{\s1\gamma}\gd{\mu\s1}\lbbBdd{A}{M}{\alpha\beta}$ \\
    & & $\hphantom{=}+ 2\Ii{A}{\mu\s1}\i{\s1\gamma}{M}\lbdd{A}{\alpha\beta}$ \\
     & $\pp{M}[,\alpha\beta]$ & $0 = \lbbdd{M}{\alpha\beta}\kd{\gamma}{\mu} - 2\kd{(\alpha}{\mu}\lbbdd{M}{\beta)\gamma} + 2\Ii{A}{\mu\s1}\gu{\s1\gamma}\lbBBdd{A}{M}{\alpha\beta} - 2\Iii{A}{\s1\gamma}\gd{\mu\s1}\lbbBBdd{A}{M}{\alpha\beta}$ \\
    & & $\hphantom{=}- 2\Iii{A}{\s1\gamma}\ii{\s1\mu}{M}\lbbdd{A}{\alpha\beta}$ \\
     & $\pp{M}[,\alpha\beta]$ & $0 = \lbbbdd{M}{\alpha\beta}\kd{\gamma}{\mu} - 2\kd{(\alpha}{\mu}\lbbbdd{M}{\beta)\gamma} + 2\Ii{A}{\mu\s1}\gu{\s1\gamma}\lbBBBdd{A}{M}{\alpha\beta} - 2\Iii{A}{\s1\gamma}\gd{\mu\s1}\lbbBBBdd{A}{M}{\alpha\beta}$ \\
    & & $\hphantom{=}+ \Iiii{A}{\mu}{\s1}\iii{\gamma}{\s1}{M}\lbbbdd{A}{\alpha\beta} - \Iiii{A}{\s1}{\gamma}\iii{\s1}{\mu}{M}\lbbbdd{A}{\alpha\beta}$ \\
    (C2) & $0$ & $0 = \Ii{A}{\mu\s1}\gu{\s1\gamma}\XX{B}{\overline{A}} - \Iii{A}{\s1\gamma}\gd{\mu\s1}\XX{B}{\overline{\overline{A}}}$ \\
    (C3) & $0$ & $0 = \Ldd{B}{\mu\nu} + 4\gu{\mu\s1}\gu{\s2\nu}\Ii{A}{\s1\s2}\TT{\overline{A}}{B} - 4\Iii{A}{\mu\nu}\TT{\overline{\overline{A}}}{B}$ \\
    (C4) & $0$ & $0 = \Ld{B}{\mu}$ \\
    & $\p{M}$ & $0 = \lbBd{M}{B}{\mu} + \Iiii{A}{\s1}{\s2}\eps{^{\mu}}{_{\s3}}{_{\s2}}\r{\i{\s1\s3}{B}\xbbbB{A}{M} + \i{\s1\s3}{M}\xbBBB{B}{A}}$ \\
     &  & $0 = \lbBBd{M}{B}{\mu} + \Iiii{A}{\s1}{\s2}\r{\eps{^\mu}{^{\s3}}{^{\s1}}\ii{\s2\s3}{B}\xbbbB{A}{M} + \eps{^\mu}{_{\s3}}{_{\s2}}\i{\s1\s3}{M}\xbbBBB{B}{A}}$ \\
     & & $0 = \lbBBBd{M}{B}{\mu} - 2\Iii{A}{\s1\s2}\eps{^\mu}{_{\s3}}{_{\s1}}\iii{\s3}{\s2}{B}\xbbB{A}{M}$ \\
     & $\pp{M}$ & $0 = \lbbBd{M}{B}{\mu} + \Iiii{A}{\s1}{\s2}\eps{^\mu}{^{\s3}}{^{\s1}}\ii{\s2\s3}{M}\xbBBB{B}{A} - \Iiii{A}{\s1}{\s2}\eps{^\mu}{_{\s2}}{_{\s3}}\i{\s1\s3}{B}\xbbbBB{A}{M}$ \\
     &  & $0 = \lbbBBd{M}{B}{\mu} + \Iiii{A}{\s1}{\s2}\eps{^\mu}{^{\s3}}{^{\s1}}\r{\ii{\s2\s3}{M}\xbbBBB{B}{A} + \ii{\s2\s3}{B}\xbbbBB{A}{M}}$ \\
     &  & $0 = \lbbBBBd{M}{B}{\mu} - 2\Ii{A}{\s1\s2}\eps{^\mu}{^{\s3}}{^{\s1}}\iii{\s2}{\s3}{B}\xbBB{A}{M}$ \\
     & $\ppp{M}$ & $0 = \lbbbBd{M}{B}{\mu} + 2\Iii{A}{\s1\s2}\eps{^\mu}{_{\s1}}{_{\s3}}\iii{\s3}{\s2}{M}\xbBB{B}{A} - \Iiii{A}{\s1}{\s2}\eps{^\mu}{_{\s2}}{_{\s3}}\i{\s1\s3}{B}\xbbbB{A}{M}$ \\
     &  & $0 = \lbbbBBd{M}{B}{\mu} + 2\Ii{A}{\s1\s2}\eps{^\mu}{^{\s1}}{^{\s3}}\iii{\s2}{\s3}{M}\xbbB{B}{A}$ \\
     & & $0 = \lbbbBBBd{M}{B}{\mu} + 2\Ii{A}{\s1\s2}\eps{^\mu}{^{\s1}}{^{\s3}}\r{\iii{\s2}{\s3}{M}\xbbbB{B}{A} + \iii{\s2}{\s3}{B}\xbBBB{A}{M}}$ \\
    & & $\hphantom{=} + 2\Iii{A}{\s1\s2}\eps{^\mu}{_{\s1}}{_{\s3}}\r{\iii{\s3}{\s2}{M}\xbbbBB{B}{A} + \iii{\s3}{\s2}{B}\xbbBBB{A}{M}}$ \\
     & $\p{M}[,\alpha]$ & $0 = 2\lbBdd{M}{B}{\mu\alpha} - \lbdBd{B}{\mu}{M}{\alpha} + \Iiii{A}{\s1}{\s2}\eudd{\mu}{\s2}{\s3}\r{\i{\s1\s3}{B}\xbbbBd{A}{M}{\alpha} + \i{\s1\s3}{M}\xbBBBd{B}{A}{\alpha}}$ \\
    & & $\hphantom{=} + 2\r{\gu{\s1\alpha}\Ii{A}{\s1\s2}\i{\s2\mu}{M} + \gu{\alpha\s1}\gu{\s2\mu}\gd{\t1\t2}\i{\t1\t2}{M}\Ii{A}{\s1\s2} - \gu{\mu\alpha}\kd{\overline{A}}{\overline{M}}}\tbB{A}{B}$ \\
    & & $\hphantom{=} + 2\r{-\gd{\s1\s2}\Iii{A}{\s1\alpha}\i{\s2\mu}{M} - \gd{\s1\s2}\i{\s1\s2}{M}\Iii{A}{\mu\alpha}}\tbBB{B}{A}$ \\
     &  & $0 = 2\lbBBdd{M}{B}{\mu\alpha} - \lbdBBd{M}{\alpha}{B}{\mu} + \Iiii{A}{\s1}{\s2}\r{\euuu{\mu}{\s1}{\s3}\ii{\s2\s3}{B}\xbbbBd{A}{M}{\alpha} + \eudd{\mu}{\s2}{\s3}\i{\s1\s3}{M}\xbbBBBd{B}{A}{\alpha}}$ \\
    & & $\hphantom{=} + 2\r{\gu{\s1\alpha}\Ii{A}{\s1\s2}\i{\s2\mu}{M} + \gu{\alpha\s1}\gu{\s2\mu}\gd{\t1\t2}\i{\t1\t2}{M}\Ii{A}{\s1\s2} - \gu{\mu\alpha}\kd{\overline{A}}{\overline{M}}}\tbBB{A}{B}$ \\
    & & $\hphantom{=} + 2\r{-\gd{\s1\s2}\Iii{A}{\s1\alpha}\i{\s2\mu}{M} - \gd{\s1\s2}\i{\s1\s2}{M}\Iii{A}{\mu\alpha}}\tbbBB{A}{B}$ \\
     &  & $0 = 2\lbBBBdd{M}{B}{\mu\alpha} - \lbdBBBd{M}{\alpha}{B}{\mu} + 2\Iii{A}{\s1\s2}\eudd{\mu}{\s3}{\s1}\iii{\s3}{\s2}{B}\xbbBd{A}{M}{\alpha}$ \\
    & & $\hphantom{=} + 2\r{\gu{\s1\alpha}\Ii{A}{\s1\s2}\i{\s2\mu}{M} + \gu{\alpha\s1}{\s2\mu}\gd{\t1\t2}\i{\t1\t2}{M}\Ii{A}{\s1\s2} - \gu{\mu\alpha}\kd{\overline{A}}{\overline{M}}}\tbBBB{A}{B}$ \\
    & & $\hphantom{=} + 2\r{-\gd{\s1\s2}\Iii{A}{\s1\alpha}\i{\s2\mu}{M} - \gd{\s1\s2}\i{\s1\s2}{M}\Iii{A}{\mu\alpha}}\tbbBBB{A}{B}$ \\
     & $\pp{M}[,\alpha]$ & $0 = 2\lbbBdd{M}{B}{\mu\alpha} - \lbdBBd{B}{\mu}{M}{\alpha} + \Iiii{A}{\s1}{\s2}\eudd{\mu}{\s2}{\s3}\i{\s1\s3}{B}\xbbbBBd{A}{M}{\alpha} - \Iiii{A}{\s1}{\s2}\euuu{\mu}{\s3}{\s1}\ii{\s2\s3}{M}\xbBBBd{B}{A}{\alpha}$ \\
    & & $\hphantom{=} + 2\r{\gu{\alpha\s1}\gu{\s2\mu}\gu{\t1\t2}\Ii{A}{\s1\s2}\ii{\t1\t2}{M} - \gu{\alpha\s1}\gu{\s2\mu}\gu{\t1\t2}\Ii{A}{\s1\t1}\ii{\s2\t2}{M}}\tbB{A}{B}$ \\
    & & $\hphantom{=} + 2\r{-\gu{\mu\alpha}\kd{\overline{\overline{A}}}{\overline{\overline{M}}} + \gu{\mu\s1}\Iii{A}{\s2\alpha}\ii{\s1\s2}{M} - \gu{\s1\s2}\Iii{A}{\mu\alpha}\ii{\s1\s2}{M}}\tbBB{B}{A}$ \\
    & & $0 = 2\lbbBBdd{M}{B}{\mu\alpha} - \lbbdBBd{B}{\mu}{M}{\alpha} + \Iiii{A}{\s1}{\s2}\euuu{\mu}{\s1}{\s3}\r{\ii{\s2\s3}{M}\xbbBBBd{B}{A}{\alpha} + \ii{\s2\s3}{B}\xbbbBBd{A}{M}{\alpha}}$ \\
    & & $\hphantom{=} + 2\r{\gu{\alpha\s1}\gu{\s2\mu}\gu{\t1\t2}\Ii{A}{\s1\s2}\ii{\t1\t2}{M} - \gu{\alpha\s1}\gu{\s2\mu}\gu{\t1\t2}\Ii{A}{\s1\t1}\ii{\s2\t2}{M}}\tbBB{A}{B}$ \\
    & & $\hphantom{=} + 2\r{-\gu{\mu\alpha}\kd{\overline{\overline{A}}}{\overline{\overline{M}}} + \gu{\mu\s1}\Iii{A}{\s2\alpha}\ii{\s1\s2}{M} - \gu{\s1\s2}\Iii{A}{\mu\alpha}\ii{\s1\s2}{M}}\tbbBB{A}{B}$ \\
     &  & $0 = 2\lbbBBBdd{M}{B}{\mu\alpha} - \lbbdBBBd{M}{\alpha}{B}{\mu} + 2\Ii{A}{\s1\s2}\euuu{\mu}{\s3}{\s1}\iii{\s2}{\s3}{B}\xbBBd{A}{M}{\alpha}$ \\
    & & $\hphantom{=} + 2\r{\gu{\alpha\s1}\gu{\s2\mu}\gu{\t1\t2}\Ii{A}{\s1\s2}\ii{\t1\t2}{M} - \gu{\alpha\s1}\gu{\s2\mu}\gu{\t1\t2}\Ii{A}{\s1\t1}\ii{\s2\t2}{M}}\tbBBB{A}{B}$ \\
    & & $\hphantom{=} + 2\r{-\gu{\mu\alpha}\kd{\overline{\overline{A}}}{\overline{\overline{M}}} + \gu{\mu\s1}\Iii{A}{\s2\alpha}\ii{\s1\s2}{M} - \gu{\s1\s2}\Iii{A}{\mu\alpha}\ii{\s1\s2}{M}}\tbbBBB{A}{B}$ \\
    & $\ppp{M}[,\alpha]$ & $0 = 2\lbbbBdd{M}{B}{\mu\alpha} - \lbdBBBd{B}{\mu}{M}{\alpha} - 2\gu{\mu\alpha}\tbBBB{B}{M}$ \\
    & & $0 = 2\lbbbBBd{M}{B}{\mu\alpha} - \lbbdBBBd{B}{\mu}{M}{\alpha} - 2\Ii{A}{\s1\s2}\euuu{\mu}{\s1}{\s3}\iii{\s2}{\s3}{M}\xbbBd{B}{A}{\alpha} - 2\gu{\mu\alpha}\tbbBBB{B}{M}$ \\
    &  & $0 = 2\lbbbBBBdd{M}{B}{\mu\alpha} - \lbbbdBBBd{B}{\mu}{M}{\alpha} + 2\Ii{A}{\s1\s2}\euuu{\mu}{\s3}{\s1}\r{\iii{\s2}{\s3}{M}\xbbbBd{B}{A}{\alpha} + \iii{\s2}{\s3}{B}\xbBBBd{A}{M}{\alpha}}$ \\
    & & $\hphantom{=} + 2\Iii{A}{\s1\s2}\eudd{\mu}{\s3}{\s1}\r{\iii{\s3}{\s2}{M}\xbbbBBd{B}{A}{\alpha} + \iii{\s3}{\s2}{B}\xbbBBBd{A}{M}{\alpha}} - 2\gu{\mu\alpha}\tbbbBBB{B}{M}$ \\
     & $\P{M}[,\alpha\beta]$ & $0 = 2\LdLdd{M}{(\alpha}{B}{\beta)\mu} - \LdLdd{B}{\mu}{M}{\alpha\beta}$ \\
    & $\P{M}[,\alpha\beta,\gamma]$ & $0 = \LddLdd{B}{\mu(\alpha}{M}{\beta\gamma)}$ \\
   	(C5) & $\p{M}$ & $0 = \Iiii{A}{\s1}{\s2}\eudd{\nu}{\s3}{\s2}\i{\s1\s3}{M}\lbbb{A}$ \\
   	 & $\pp{M}$ & $0 = \Iiii{A}{\s1}{\s2}\euuu{\nu}{\s3}{\s1}\ii{\s2\s3}{M}\lbbb{A}$ \\
   	 & $\ppp{M}$ & $0 = \Ii{A}{\s1\s2}\euuu{\nu}{\s1}{\s3}\iii{\s2}{\s3}{M}\lb{A} + \Iii{A}{\s1\s2}\eudd{\nu}{\s1}{\s3}\iii{\s3}{\s2}{M}\lbb{A}$ \\
   	 & $\P{M}[,\alpha]$ & $0 = \gu{\alpha\s1}\gu{\s2\nu}\Ii{A}{\s1\s2}\XX{\overline{A}}{M} - \Iii{A}{\alpha\nu}\XX{\overline{\overline{A}}}{M}$ \\
   	 & $\p{M}[,\alpha\beta]$ & $0 = -4\Iii{A}{\nu(\alpha}\xbbBd{A}{M}{\beta)} + \Iiii{A}{\s1}{\s2}\eudd{\nu}{\s3}{s2}\i{\s1\s3}{M}\lbbbdd{A}{\alpha\beta}$ \\
   	 & $\pp{M}[,\alpha\beta]$ & $0 = 4\gu{\nu\s1}\gu{\s2(\alpha}\Ii{A}{\s1\s2}\xbBBd{A}{M}{\beta)} + \Iiii{A}{\s1}{\s2}\euuu{\nu}{\s3}{\s1}\ii{\s2\s3}{M}\lbbbdd{A}{\alpha\beta}$ \\
    & $\ppp{M}[,\alpha\beta]$ & $0 = 4\r{\gu{\nu\s1}\gu{\s2(\alpha}\Ii{A}{\s1\s2}\xbBBBd{A}{M}{\beta)} - \Iii{A}{\nu(\alpha}\xbbBBBd{A}{M}{\beta)}}$ \\
   	& & $\hphantom{=} + 2\r{\Ii{A}{\s1\s2}\euuu{\nu}{\s1}{\s3}\iii{\s2}{\s3}{M}\lbdd{A}{\alpha\beta} + \Iii{A}{\s1\s2}\eudd{\nu}{\s1}{\s3}\iii{\s3}{\s2}{M}\lbbdd{A}{\alpha\beta}}$ \\
   	 & $\p{M}\p{N}$ & $0 = \Iiii{A}{\s1}{\s2}\eudd{\nu}{\s3}{\s2}\r{\i{\s1\s3}{M}\lbBBB{N}{A} + \i{\s1\s3}{N}\lbBBB{M}{A}}$ \\
   	 & $\p{M}\pp{N}$ & $0 = \Iiii{A}{\s1}{\s2}\eudd{\nu}{\s3}{\s2}\i{\s1\s3}{M}\lbbBBB{N}{A} + \Iiii{A}{\s1}{\s2}\euuu{\nu}{\s3}{\s1}\ii{\s2\s3}{N}\lbBBB{M}{A}$ \\
   	 & $\p{M}\ppp{N}$ & $0 = \Iiii{A}{\s1}{\s2}\eudd{\nu}{\s3}{\s2}\i{\s1\s3}{M}\lbbbBBB{A}{N} + 2\r{\Ii{A}{\s1\s2}\euuu{\nu}{\s1}{\s3}\iii{\s2}{\s3}{N}\lbB{A}{M} + \Iii{A}{\s1\s2}\eudd{\nu}{\s1}{\s3}\iii{\s3}{\s2}{N}\lbBB{M}{A}}$ \\
   	& & $\hphantom{=} + \Ii{A}{\s1\s2}\euuu{\nu}{\s1}{\s3}\r{\gd{\t1\t2}\i{\t1\t2}{M}\iii{\s2}{\s3}{N} + \gd{\t1\t2}\i{\t1\s2}{M}\iii{\t2}{\s3}{N} - \gd{\t1\s3}\i{\t1\t2}{M}\iii{\s2}{\t2}{N}}\lb{A}$ \\
   	& & $\hphantom{=} + \Iii{A}{\s1\s2}\eddd{\s3}{\s4}{\s1}\r{\gu{\s3\nu}\gd{\t1\s2}\i{\t1\t2}{M}\iii{\s4}{\t2}{N} - \gu{\s3\nu}\gd{\t1\t2}\i{\t1\s4}{M}\iii{\t2}{\s2}{N} - \i{\s3\nu}{M}\iii{\s4}{\s2}{N}}\lbb{A}$ \\
 	 & $\pp{M}\pp{N}$ & $0 = \Iiii{A}{\s1}{\s2}\euuu{\nu}{\s3}{\s1}\r{\ii{\s2\s3}{M}\lbbBBB{N}{A} + \ii{\s2\s3}{N}\lbbBBB{M}{A}}$ \\
 	 & $\pp{M}\ppp{N}$ & $0 = \Iiii{A}{\s1}{\s2}\euuu{\nu}{\s3}{\s1}\ii{\s2\s3}{M}\lbbBBB{A}{N} + 2\r{\Ii{A}{\s1\s2}\euuu{\nu}{\s1}{\s3}\iii{\s2}{\s3}{N}\lbBB{A}{M} + \Iii{A}{\s1\s2}\eudd{\nu}{\s1}{\s3}\iii{\s3}{\s2}{N}\lbbBB{A}{M}}$ \\
 	& & $\hphantom{=} + \Iii{A}{\s1\s2}\eddd{\s3}{\s4}{\s1}\r{\gu{\s3\t1}\gu{\t2\nu}\ii{\t1\t2}{M}\iii{\s4}{\s2}{N} - \gu{\s3\nu}\gu{\t1\t2}\ii{\t1\t2}{M}\iii{\s4}{\s2}{N}}\lbb{A}$ \\
 	 & $\ppp{M}\ppp{N}$ & $0 = \Ii{A}{\s1\s2}\euuu{\nu}{\s1}{\s3}\r{\iii{\s2}{\s3}{M}\lbBBB{A}{N} + \iii{\s2}{\s3}{N}{\lbBBB{A}{M}}}$ \\
 	& & $\hphantom{=} + \Iii{A}{\s1\s2}\eudd{\nu}{\s1}{\s3}\r{\iii{\s3}{\s2}{M}\lbbBBB{A}{N} + \iii{\s3}{\s2}{N}\lbbBBB{A}{M}}$ \\
 	(C6) & $0$ & $0 = \Iiii{A}{\s1}{\s2}\eudd{\gamma}{\s2}{\s3}\r{\i{\s1\s3}{M}\tbBBB{B}{A} + \i{\s1\s3}{B}\tbBBB{M}{A}}$ \\
 	 & & $0 = 2\Iiii{A}{\s1}{\s2}\eudd{\gamma}{\s2}{\s3}\i{\s1\s3}{M}\tbbBBB{B}{A} + 2\Iiii{A}{\s1}{\s2}\euuu{\gamma}{\s1}{\s3}\ii{\s2\s3}{B}\tbBBB{M}{A} - \xbbBd{B}{M}{\gamma} - \xbBBd{M}{B}{\gamma}$ \\
 	 &  & $0 = 2\Iiii{A}{\s1}{\s2}\eudd{\gamma}{\s2}{\s3}\i{\s1\s3}{M}\tbbbBBB{A}{B} + 4\Ii{A}{\s1\s2}\euuu{\gamma}{\s3}{\s1}\iii{\s2}{\s3}{B}\tbB{A}{M}$ \\
 	& & $\hphantom{=} + 4\Iii{A}{\s1\s2}\eudd{\gamma}{\s3}{\s1}\iii{\s3}{\s2}{B}\tbBB{M}{A} - \xbbbBd{B}{M}{\gamma} - \xbBBBd{M}{B}{\gamma}$ \\
 	 &  & $0 = \Iiii{A}{\s1}{\s2}\euuu{\gamma}{\s1}{\s3}\r{\ii{\s2\s3}{B}\tbbBBB{M}{A} + \ii{\s2\s3}{M}\tbbBBB{B}{A}}$ \\
 	 &  & $0 = 2\Iiii{A}{\s1}{\s2}\euuu{\gamma}{\s1}{\s3}\ii{\s2\s3}{M}\tbbbBBB{A}{B} + 4 \Ii{A}{\s1\s2}\euuu{\gamma}{\s3}{\s1}\iii{\s2}{\s3}{B}\tbBB{A}{M}$ \\
 	& & $\hphantom{=} + 4\Iii{A}{\s1\s2}\eudd{\gamma}{\s3}{\s1}\iii{\s3}{\s2}{B}\tbbBB{A}{M} - \xbbbBBd{B}{M}{\gamma} - \xbbBBBd{M}{B}{\gamma}$ \\
     &  & $0 = \Ii{A}{\s1\s2}\euuu{\gamma}{\s3}{\s1}\r{\iii{\s2}{\s3}{M}\tbBBB{A}{B} + \iii{\s2}{\s3}{B}\tbBBB{A}{M}}$ \\
 	& & $\hphantom{=} + \Iii{A}{\s1\s2}\eudd{\gamma}{\s3}{\s1}\r{\iii{\s3}{\s2}{M}\tbbBBB{A}{B} + \iii{\s3}{\s2}{B}\tbbBBB{A}{M}}$ \\
 	(C$\text{8}_\text{2}$) & $0$ & $0 = \Ii{A}{\mu\s1}\gu{\s1(\b1}\lbd{A}{\b2)} - \gd{\mu\s1}\Iii{A}{\s1(\b1}\lbbd{A}{\b2)}$ \\
 	 & $\P{M}$ & $0 = \Ii{A}{\mu\s1}\gu{\s1(\b1}\LLd{M}{\overline{A}}{\b2)} - \gd{\mu\s1}\Iii{A}{\s1(\b1}\LLd{M}{\overline{\overline{A}}}{\b2)}$ \\
 	 & $\p{M}[,\alpha]$ & $0 = \kd{\alpha}{\mu}\lbdd{M}{\b1\b2} - 2\Ii{A}{\mu\s1}\gu{\s1(\b1}\lbdBd{A}{\b2)}{M}{\alpha} + 2\Iii{A}{\s1(\b1|}\gd{\mu\s1}\lbdBBd{M}{\alpha}{A}{|\b2)} - 4\Ii{A}{\mu\s1}\i{\s1(\b1}{M}\lbdd{A}{\b2)\alpha}$ \\
 	 & $\pp{M}[,\alpha]$ & $0 = \kd{\alpha}{\mu}\lbbdd{M}{\b1\b2} - 2\Ii{A}{\mu\s1}\gu{\s1(\b1}\lbdBBd{A}{\b2)}{M}{\alpha} + 2\gd{\mu\s1}\Iii{A}{\s1(\b1}\lbbdBBd{A}{\b2)}{M}{\alpha} + 4\Iii{A}{\s1(\b1|}\ii{\mu\s1}{M}\lbbdd{A}{|\b2)\alpha}$ \\
 	 & $\ppp{M}[,\alpha]$ & $0 = \kd{\alpha}{\mu}\lbbbdd{M}{\b1\b2} - 2\Ii{A}{\mu\s1}\gu{\s1(\b1}\lbdBBBd{A}{\b2)}{M}{\alpha} + 2\Iii{A}{\s1(\b1|}\gd{\mu\s1}\lbbdBBBd{A}{|\b2)}{M}{\alpha} - 2\Iiii{A}{\mu}{\s1}\iii{(\b1|}{\s1}{M}\lbbbdd{A}{|\b2)\alpha}$ \\
 	& & $\hphantom{=} + 2\iii{\s1}{\mu}{M}\Iiii{A}{(\b1}{\s1}\lbbbdd{A}{\b2)\alpha}$ \\
 	 & $\P{M}[,\alpha\beta]$ & $0 = \Ii{A}{\mu\s1}\gu{\s1(\b1}\LdLdd{\overline{A}}{\b2)}{M}{\alpha\beta} - \Iii{A}{\s1(\b1|}\gd{\mu\s1}\LdLdd{A}{|\b2)}{M}{\alpha\beta}$ \\
 	(C$\text{8}_\text{3}$) & $0$ & $0 = \Ii{A}{\mu\s1}\gu{\s1(\b1}\lbdd{A}{\b2\b3)} - \Iii{A}{\s1(\b1|}\gd{\mu\s1}\lbbdd{A}{|\b2\b3)}$ \\
 	 & $\p{M}$ & $0 = \Ii{A}{\mu\s1}\gu{\s1(\b1}\lbBdd{M}{A}{\b2\b3)} - \Iii{A}{\s1(\b1|}\gd{\mu\s1}\lbBBdd{M}{A}{|\b2\b3)} + \Ii{A}{\mu\s1}\i{\s1(\b1}{M}\lbdd{A}{\b2\b3)}$ \\
 	 & $\pp{M}$ & $0 = \Ii{A}{\mu\s1}\gu{\s1(\b1}\lbbBdd{M}{A}{\b2\b3)} - \Iii{A}{\s1(\b1|}\gd{\mu\s1}\lbbBBdd{M}{A}{|\b2\b3)} - \Iii{A}{\s1(\b1|}\ii{\mu\s1}{M}\lbbdd{A}{|\b2\b3)}$ \\
 	 & $\ppp{M}$ & $0 = 2\Ii{A}{\mu\s1}\gu{\s1(\b1}\lbbbBdd{M}{A}{\b2\b3)} - 2\Iii{A}{\s1(\b1|}\gd{\mu\s1}\lbbbBBdd{M}{A}{|\b2\b3)} + \Iiii{A}{\mu}{\s1}\iii{(\b1}{\s1}{M}\lbbbdd{A}{\b2\b3)} $\\
 	& & $\hphantom{=}- \iii{\s1}{\mu}{M}\Iiii{A}{\s1}{(\b1}\lbbbdd{A}{\b2\b3)}$ \\
 	 & $\P{M}[,\alpha]$ & $0 = \Ii{A}{\mu\s1}\gu{\s1(\b1|}\LdLdd{M}{\alpha}{\overline{A}}{|\b2\b3)} - \Iii{A}{\s1(\b1|}\gd{\mu\s1}\LdLdd{M}{\alpha}{\overline{\overline{A}}}{|\b2\b3)}$ \\
 	 & $\P{M}[,\alpha\beta]$ & $0 = \Ii{A}{\mu\s1}\gu{\s1(\b1|}\LddLdd{M}{\alpha\beta}{\overline{A}}{|\b2\b3)} - \Iii{A}{\s1(\b1|}\gd{\mu\s1}\LddLdd{M}{\alpha\beta}{\overline{\overline{A}}}{|\b2\b3)}$ \\
 	(C$\text{9}_\text{2}$) & $0$ & $0 = \Ii{A}{\mu\s1}\gu{\s1(\b1}\XXd{B}{\overline{A}}{\b2)} - \Iii{A}{\s1(\b1|}\gd{\mu\s1}\XXd{B}{\overline{\overline{A}}}{\b2)}$ \\
 	(C$\text{21}_\text{3}$) & $\p{M}$ & $0 = \Iiii{A}{\s1}{\s2}\i{\s1\s3}{M}\eudd{(\m1}{\s3}{\s2}\lbbbdd{A}{\m2\m3)}$ \\
 	 & $\pp{M}$ & $0 = \Iiii{A}{\s1}{\s2}\ii{\s2\s3}{M}\euuu{\s3}{\s1}{(\m1}\lbbbdd{A}{\m2\m3)}$ \\
 	 & $\ppp{M}$ & $0 = \Ii{A}{\s1\s2}\iii{\s2}{\s3}{M}\euuu{\s1}{\s3}{(\m1}\lbdd{A}{\m2\m3)} + \Iii{A}{\s1\s2}\iii{\s3}{\s2}{M}\eudd{(\m1}{\s1}{\s3}\lbbdd{A}{\m2\m3)}$ \\
 	
 	 & $\p{M}\p{N}$ & $0 = \frac{1}{2}\Iiii{A}{\s1}{\s2}\r{\gd{\s4\s5}\i{\s3\s5}{M}\i{\s1\s4}{N} + \gd{\s4\s5}\i{\s3\s5}{N}\i{\s1\s4}{M}}\eudd{(\m1|}{\s2}{\s3}\lbbbdd{A}{|\m2\m3)}$ \\
 	& & $\hphantom{=} + \frac{1}{2}\Iiii{A}{\s1}{\s2}\r{\gd{\s3\s4}\i{\s3\s4}{M}\i{\s1\s5}{N} + \gd{\s3\s4}\i{\s3\s4}{N}\i{\s1\s5}{M}}\eudd{(\m1|}{\s5}{\s2}\lbbbdd{A}{|\m2\m3)}$ \\
 	& & $\hphantom{=} + \Iiii{A}{\s1}{\s2}\eudd{(\m1|}{\s3}{\s2}\r{\i{\s1\s3}{M}\lbBBBdd{N}{A}{|\m2\m3)} + \i{\s1\s3}{N}\lbBBBdd{M}{A}{|\m2\m3)}}$ \\
 	 & $\p{M}\pp{N}$ & $0 = \frac{1}{2}\Iiii{A}{\s1}{\s2}\rp{\i{\s1\s4}{M}\ii{\s3\s4}{N}\euud{(\m1|}{\s3}{\s2} + \i{\s1\s4}{M}\ii{\s2\s3}{N}\eudu{(\m1|}{\s4}{\s3}}$ \\
 	& & $\hphantom{=} + \pr{\gd{\s4\s5}\i{\s4\s5}{M}\ii{\s2\s3}{N}\euuu{\s3}{\s1}{(\m1|}}\lbbbdd{A}{|\m2\m3)} + \Iiii{A}{\s1}{\s2}\i{\s1\s3}{M}\eudd{(\m1|}{\s3}{\s2}\lbbBBBdd{N}{A}{|\m2\m3)}$ \\
 	& & $\hphantom{=} + \Iiii{A}{\s1}{\s2}\ii{\s2\s3}{N}\euuu{\s3}{\s1}{(\m1|}\lbBBBdd{M}{A}{|\m2\m3)}$ \\
 	 & $\p{M}\ppp{N}$ & $0 = \Ii{A}{\s1\s2}\r{\gd{\t1\t2}\i{\t1\t2}{M}\iii{\s2}{\s3}{N} + \gd{\t1\t2}\i{\t1\s2}{M}\iii{\t2}{\s3}{N} - \gd{\t1\s3}\i{\t1\t2}{M}\iii{\s2}{\t2}{N}}\euuu{\s1}{\s3}{(\m1}\lbdd{A}{\m2\m3)}$ \\
 	& & $\hphantom{=} + \Iii{A}{\s1\s2}\eddd{\s3}{\s4}{\s1}\rp{\i{\t1\t2}{M}\iii{\s4}{\t2}{N}\gd{\t1\s2}\gu{\s3(\m1|} - \i{\t1\s4}{M}\iii{\t2}{\s2}{N}\gd{\t1\t2}\gu{\s3(\m1|}}$ \\
 	& & $\hphantom{=} - \pr{\iii{\s4}{\s2}{N}\i{\s3(\m1|}{M}}\lbbdd{A}{|\m2\m3)} + \Iiii{A}{\s1}{\s2}\i{\s1\s3}{M}\eudd{(\m1|}{\s3}{\s2}\lbbbBBBdd{N}{A}{|\m2\m3)}$ \\
 	& & $\hphantom{=} + 2\Ii{A}{\s1\s2}\iii{\s2}{\s3}{N}\euuu{\s1}{\s3}{(\m1}\lbBdd{M}{A}{\m2\m3)} + 2\Iii{A}{\s1\s2}\iii{\s3}{\s2}{N}\eudd{(\m1|}{\s1}{\s3}\lbBBdd{M}{A}{|\m2\m3)}$ \\
 	 & $\pp{M}\pp{N}$ & $0 = \Iiii{A}{\s1}{\s2}\euuu{\s3}{\s1}{(\m1|}\r{\ii{\s3\s2}{M}\lbbBBBdd{N}{A}{|\m2\m3)} + \ii{\s3\s2}{N}\lbbBBBdd{M}{A}{|\m2\m3)}}$ \\
 	 & $\pp{M}\ppp{N}$ & $0 = \Iii{A}{\s1\s2}\eddd{\s3}{\s4}{\s1}\r{\ii{\t1\t2}{M}\iii{\s4}{\s2}{N}\gu{\s3\t1}\gu{\t2(\m1|} - \ii{\t1\t2}{M}\iii{\s4}{\s2}{N}\gu{\t1\t2}\gu{\s3(\m1|}}\lbbdd{A}{|\m2\m3)}$ \\
 	& & $\hphantom{=} + \Iiii{A}{\s1}{\s2}\ii{\s2\s3}{M}\euuu{\s3}{\s1}{(\m1|}\lbbbBBBdd{N}{A}{|\m2\m3)} + 2\Ii{A}{\s1\s2}\iii{\s2}{\s3}{N}\euuu{\s1}{\s3}{(\m1}\lbbBdd{M}{A}{\m2\m3)}$ \\
 	& & $\hphantom{=} + 2\Iii{A}{\s1\s2}\iii{\s3}{\s2}{N}\eudd{(\m1|}{\s1}{\s3}\lbbBBdd{M}{A}{|\m2\m3)}$ \\
 	 & $\ppp{M}\ppp{N}$ & $0 = \Ii{A}{\s1\s2}\euuu{\s1}{\s3}{(\m1}\r{\iii{\s2}{\s3}{M}\lbbbBdd{N}{A}{\m2\m3)} + \iii{\s2}{\s3}{N}\lbbbBdd{M}{A}{\m2\m3)}}$ \\
 	& & $\hphantom{=} + \Iii{A}{\s1\s2}\eudd{(\m1|}{\s1}{\s3}\r{\iii{\s3}{\s2}{M}\lbbbBBdd{N}{A}{|\m2\m3)} + \iii{\s3}{\s2}{N}\lbbbBBdd{M}{A}{|\m2\m3)}}$ \\
 	& $\p{M}\P{N}[,\alpha]$ & $0 = \Iiii{A}{\s1}{\s2}\i{\s1\s3}{M}\eudd{(\m1|}{\s3}{\s2}\LdLdd{N}{\alpha}{\overline{\overline{\overline{A}}}}{|\m2\m3)}$ \\
 	 & $\p{M}\P{N}[,\alpha\beta]$ & $0 = \Iiii{A}{\s1}{\s2}\i{\s1\s3}{M}\eudd{(\m1|}{\s3}{\s2}\LddLdd{N}{\alpha\beta}{\overline{\overline{\overline{A}}}}{|\m2\m3)}$ \\
 	& $\pp{M}\P{N}[,\alpha]$ & $0 = \Iiii{A}{\s1}{\s2}\ii{\s2\s3}{M}\euuu{\s3}{\s1}{(\m1|}\LdLdd{N}{\alpha}{\overline{\overline{\overline{A}}}}{|\m2\m3)}$ \\
 	 & $\pp{M}\P{N}[,\alpha\beta]$ & $0 = \Iiii{A}{\s1}{\s2}\ii{\s2\s3}{M}\euuu{\s3}{\s1}{(\m1|}\LddLdd{N}{\alpha\beta}{\overline{\overline{\overline{A}}}}{|\m2\m3)}$ \\
  & $\ppp{M}\P{N}[,\alpha]$ & $0 = \Ii{A}{\s1\s2}\iii{\s2}{\s3}{M}\euuu{\s1}{\s3}{(\m1|}\LdLdd{N}{\alpha}{\overline{A}}{|\m2\m3)} + \Iii{A}{\s1\s2}\iii{\s3}{\s2}{M}\eudd{(\m1|}{\s1}{\s3}\LdLdd{N}{\alpha}{\overline{\overline{A}}}{|\m2\m3)}$ \\
 	 & $\ppp{M}\P{N}[,\alpha\beta]$ & $0 = \Ii{A}{\s1\s2}\iii{\s2}{\s3}{M}\euuu{\s1}{\s3}{(\m1|}\LddLdd{N}{\alpha\beta}{\overline{A}}{|\m2\m3)} + \Iii{A}{\s1\s2}\iii{\s3}{\s2}{M}\eudd{(\m1|}{\s1}{\s3}\LddLdd{N}{\alpha\beta}{\overline{\overline{A}}}{|\m2\m3)}$ \\[6pt]
 	\hline
\end{longtable}

 \end{document}